\documentclass[structabstract]{aa} 
\usepackage{graphicx}
\usepackage{natbib}
\usepackage{txfonts}
\bibpunct{(}{)}{;}{a}{}{,} 	% follow A&A style

\begin{document}

   \title{Three-dimensional surface convection simulations of \\ metal-poor stars}

   \subtitle{The effect of scattering on the photospheric temperature stratification}

   \author{	R. Collet \inst{1} \and
   			%\fnmsep\thanks{}
			W. Hayek \inst{2,1,3}  \and
			M. Asplund \inst{1} \and
			{\AA}. Nordlund \inst{4} \and
			R. Trampedach \inst{5} \and
			B. Gudiksen \inst{3}
		  }

   \institute{ Max-Planck-Institut f{\"u}r Astrophysik, Karl-Schwarzschild-Str.~1, D--85741 Garching b. M{\"u}nchen, Germany \\
   				\email{remo,hayek,asplund@mpa-garching.mpg.de}
	          %\thanks{}
         \and
             Research School of Astronomy \& Astrophysics, Cotter Road, Weston ACT 2611, Australia 
         \and
             Institute for Theoretical Astrophysics, Sem S{\ae}lands Vei, N--0315 Oslo, Norway
             \email{b.v.gudiksen@astro.uio.no}
         \and
         	 Niels Bohr Institute, University of Copenhagen, Juliane Maries Vej 30, DK--2100, Copenhagen, Denmark
	 \email{aake@nbi.dk}
         \and
             JILA, University of Colorado, 440 UCB, Boulder, CO 80309-0440, USA
             \email{trampeda@lcd.colorado.edu}
             }

   \offprints{R. Collet, \\ \email{remo@mpa-garching.mpg.de}}

   %%\date{}
   \date{Received November 15, 2010; accepted January 8, 2011}
 
  \abstract
   {Three-dimensional (3D) radiative hydrodynamic model atmospheres of metal-poor late-type stars are characterized by cooler upper photospheric layers than their one-dimensional counterparts. This property of 3D model atmospheres can dramatically affect the determination of elemental abundances from temperature-sensitive spectral features, with profound consequences on galactic chemical evolution studies.}
   {We investigate whether the cool surface temperatures predicted by 3D model atmospheres of metal-poor stars can be ascribed to approximations in the treatment of scattering during the modelling phase.}
   {We use the \textsc{Bifrost} code to construct 3D model atmospheres of metal-poor stars and test three different ways to handle scattering in the radiative transfer equation.
    As a first approach, we solve iteratively the radiative transfer equation for the general case of a source function with a coherent scattering term, treating scattering in a correct and consistent way.
    As a second approach, we solve the radiative transfer equation in local thermodynamic equilibrium approximation, neglecting altogether the contribution of continuum scattering to extinction in the optically thin layers; this has been the default mode in our previous 3D modelling as well as in present \textsc{Stagger-Code} models.
    As our third and final approach, we treat continuum scattering as pure absorption everywhere, which is the standard case in the 3D modelling by the \textsc{CO$^5$BOLD} collaboration.}
   {For all simulations, we find that the second approach produces temperature structures with cool upper photospheric layers very similar to the case in which scattering is treated correctly. 
   In contrast, treating scattering as pure absorption leads instead to significantly hotter and shallower temperature stratifications.
   The main differences in temperature structure between our published models computed with the \textsc{Stagger-} and \textsc{Bifrost} codes and those generated with the \textsc{CO$^5$BOLD} code can be traced to the different treatments of scattering.}
   {Neglecting the contribution of continuum scattering to extinction in optically thin layers provides a good approximation to the full, iterative solution of the radiative transfer equation in metal-poor stellar surface convection simulations, and at a much lower computational cost.
 Our results also demonstrate that the cool temperature stratifications predicted for metal-poor late-type stars by previous models by our collaboration are not an artifact of the approximated treatment of scattering.}


   \maketitle
%
%________________________________________________________________

\section{Introduction}

Model stellar atmospheres are indispensable tools for the quantitative interpretation of observed stellar spectra and colours.
Synthetic stellar fluxes from model stellar atmospheres can for instance be compared with observations to infer the physical properties and chemical compositions of distant stars.

The vast majority of theoretical model atmospheres of solar- and late-type stars in use today are still computed under the assumption of one-dimensional (1D) geometry, flux constancy, and hydrostatic equilibrium. 
Yet, convective energy transport -- which in late-type stars significantly affects the photospheric temperature stratification -- can be modelled in 1D only by means of approximate descriptions such as the mixing-length theory (MLT) by \citet{boehm-vitense58} or the full spectrum of turbulence (FST) method by \citet{canuto91}, which are all hampered by a number of free parameters.

Recent years, on the other hand, have seen a rapid development of numerical three-dimensional (3D) radiative hydrodynamic simulations of stellar surface convection. In these simulations, the gas flows in the highly stratified outer layers of stars are modelled by solving the hydrodynamics equations of mass, momentum, and energy conservation and accounting for the energy exchanges between matter and radiation via radiative transfer.
In this framework, convective motions arise and self-organize naturally without the need to introduce adjustable parameters.
One of the main goals of 3D modelling of stellar surface convection is therefore to provide a more realistic description of the physical structure of late-type stellar atmospheres.
Since the early studies by, e.g., \citet{nordlund82}, \citet{nordlund90}, and \citet{steffen89}, convection simulations have become increasingly more sophisticated not only from the point of view of the implemented numerical algorithms, but also in terms of input physics, allowing for quantitative comparisons with observations. 
Surface convection simulations have been successfully applied to studying the properties of granulation at the surface of the Sun \citep[e.g.][]{stein98,carlsson04,danilovic08,pereira09a,nordlund09,wedemeyer09} and other late-type stars \citep[e.g.][]{allende02a,ramirez09,chiavassa10c}. 
Theoretical predictions of photospheric temperature stratifications from 3D model stellar atmospheres have been tested in a number of cases against observed centre-to-limb variations, showing in general excellent agreement with the measurements \cite[][]{aufdenberg05,bigot06,koesterke08,pereira09b,chiavassa10c}.

Though not yet commonplace, 3D model atmospheres have also started to be employed in spectroscopic analyses for the determination of stellar elemental abundances \citep[e.g.][]{asplund99,asplund03,agss09,asplund01,collet06,collet07,caffau10,behara10,gonzalez10}.
The temperature and density inhomogeneities and the velocity fields present in 3D hydrodynamic simulations but absent in 1D hydrostatic models are in general responsible for differences between 3D and 1D analyses in terms of predicted shapes and strengths of synthetic spectral lines and in terms of abundances inferred from observed stellar spectra.
For very metal-poor stars, in addition, another aspect contributes to differences between 3D and 1D spetroscopic analyses.
Three-dimensional surface convection simulations predict in fact much cooler temperature stratifications with steeper gradients than 1D hydrostatic models in the upper layers of very metal-poor late-type stellar atmospheres \citep{asplund99,collet07}.
In the upper layers of 3D hydrodynamic simulations, the temperature stratification is determined by the competition between radiative heating and adiabatic cooling owing to the expansion of gas overshooting into the atmosphere from the top of the convection zone.
In solar-metallicity 3D simulations, spectral lines contribute enough heating in the upper atmosphere to maintain the temperature stratification close to a purely radiative equilibrium one, as in 1D models. 
At very low metallicities, on the contrary, the scarcity and overall weakness of line opacity sources and the cooling effect owing to the expansion of gas above granules act together to keep the average temperature low in the upper photosphere.
In corresponding stationary 1D hydrostatic models, the cooling component associated with gas flows is missing and balance is reached at higher temperatures.
From the point of view of spectroscopic analyses, these structural differences between 3D and 1D models can lead to large differences (up to about $1$~dex) in terms of derived abundances whenever temperature-sensitive spectral features are used \citep{asplund05}.
While there is general qualitative agreement on the above results, the predicted average equilibrium temperature stratifications of 3D model metal-poor atmospheres from different groups often vary significantly depending on the adopted codes and input physics.
These differences are partly responsible for discrepancies in terms of computed 3D$-$1D corrections to elemental abundances from different studies \citep{bonifacio10,dobrovolskas10,ivanauskas10}.

We propose that different treatments of scattering in the solution of the radiative transfer equation are responsible for the differences among various convection codes in terms of predicted photospheric temperature stratifications.
We explore for the first time the effect of coherent isotropic continuum scattering on the energy balance in the upper layers of 3D stellar surface convection simulations of metal-poor late-type stars. 
We compute simulations where we iteratively solve the radiative transfer problem for a source function that includes a coherent scattering term, and compare them with other simulations where scattering is treated by means of two commonly adopted approximations, by either neglecting it altogether in the optically thin layers or by including it as true absorption everywhere.

%__________________________________________________________________

\section{Methods}
\label{sec:methods}

\subsection{Surface convection simulations}
\label{sec:simulations}
We carried out surface convection simulations of two metal-poor red giants and one metal-poor turnoff star using the 3D, time-dependent, radiation-hydrodynamic {\sc Bifrost} code (\citealt{hayek10}; \citealt{carlsson10}; Gudiksen et al., in prep.).
The mass, momentum, and energy conservation equations are discretized using a 
high-order finite-difference scheme (6$^\mathrm{th}$ order derivatives, 5$^\mathrm{th}$ order interpolations) and solved on a staggered Eulerian mesh for a representative, 3D, rectangular, volume located across the optical surface. 
We employ a numerical resolution of  $240^3$ for the mesh, with five  ghost zone layers at the top and the bottom of the simulation box.
As for the boundary conditions, we adopt periodic boundary conditions horizontally and open boundaries vertically. 
At the bottom of the simulation domain, the inflowing material is set to have constant entropy and pressure, while the outflowing material is free to carry entropy fluctuations out.
In the last physical layer at the top of the simulation box, the internal energy per unit mass is forced not to deviate by more than 10\% from its average value, for stability reasons. 
Also, for inflowing gas, the internal energy per unit mass is then set to its minimum value from the uppermost five physical layers at the top.

The stellar parameters of the simulations are summarized in Table~\ref{tab:param}. For each simulation, we assume a solar composition \citep{ags05} with the abundances of all metals scaled proportionally to the iron abundance. 
We chose the physical dimensions of the simulations to be sufficiently large to allow about ten major granules to develop at any one time in the domain and to span about twelve pressure scale heights vertically.
In terms of Rosseland optical depth, the simulations cover the range ${-4}{\la}\log\tau_{\mathrm{Ross}}{\la}{6}$.
As initial conditions, we used snapshots from previous simulation sequences by \citet{collet07} and \citet{asplund01} and converted them to the desired numerical resolution.
We then let the simulations achieve complete thermal relaxation by running them over a period of one or two convective overturn times.\footnote{We define here the simulation's \emph{convective overturn time} as the characteristic time it takes for a fluid parcel to travel through the entire depth of the simulation domain, from the bottom to the surface and then back.} 
Given that the fundamental stellar parameters of the simulations were not varied and that the changes owing to updates of the input micro-physics are relatively small, relaxation is rapidly achieved during this time span.

\subsection{Radiative transfer}
An accurate representation of radiative energy exchange is essential for a realistic description of the temperature stratification in stellar surface layers. 
We describe the interaction between the plasma and radiation via a radiative heating term in the energy conservation equation:
\begin{equation}
Q_\mathrm{rad} = \int_\Omega \! \int_0^\infty \!  \chi_\lambda \, (I_\lambda-S_{\!\lambda}) \, d\lambda~d\Omega,
\label{eq:qradmon}
\end{equation}
where $\chi_\lambda$ is the total monochromatic extinction coefficient (per unit path length), sum of the absorption $\kappa_\lambda$ and scattering $\sigma_\lambda$ extinction coefficients, $S_{\!\lambda}$ and $I_\lambda$ are the monochromatic source function and intensity, respectively, and the integral is performed over the whole wavelength range and over the whole solid angle. 
Under the assumption of an isotropic source function and extinction coefficients and of a static medium, integral~(\ref{eq:qradmon}) is equivalent to
\begin{equation}
Q_\mathrm{rad} = 4 \pi \! \int_0^\infty \!  \chi_\lambda \, (J_\lambda-S_{\!\lambda}) \, d\lambda,
\label{eq:qradmonj}
\end{equation}
where $J_\lambda\!=\!\frac{1}{4\pi}\int_\Omega I_\lambda\,d\Omega$ is the monochromatic mean intensity.

We adopt a realistic equation of state \citep{mihalas88}, which accounts for the effects of excitation, ionization, and dissociation of the 15 most abundant elements and of $H_2$ and $H_2^+$, continuous opacities from \citet{gustafsson75} and Trampedach (2011, in prep.), and sampled line opacities from B. Plez \citep[priv. comm.; see also][]{gustafsson08}. The full list of references for the opacity sources is given in \citet{hayek10}.
In our simulations, we compute the radiative heating rates by solving the radiative transfer equation at each time step and at each grid-point in three-dimensions on short characteristics along 24~rays.
The integral over solid angle in Eq.~\ref{eq:qradmon} is then approximated by a weighted Carlson's A4 quadrature sum \citep{carlson63} of the $I_\lambda-S_{\!\lambda}$ terms at these specific directions.
The implemented radiative transfer method \citep{hayek10} allows for a solution of the radiative transfer equation with coherent scattering by means of an iterative Gauss-Seidel acceleration scheme \citep{trujillo95}.
In the latter case, the monochromatic source function is expressed in the form $S_{\!\lambda}=\epsilon_\lambda {B}_\lambda+(1\!-\!\epsilon_\lambda){J_\lambda}$, where $\epsilon_\lambda{\equiv}\kappa_\lambda/\chi_\lambda$ is the photon destruction probability, $B_\lambda$ is the Planck function, and ${\epsilon_\lambda}{B_\lambda}$ is the thermal emissivity. 
Below, we will refer to the solution of the radiative transfer equation for a source function with a coherent scattering term as the standard case.
With a source function of the above form, it is easy to show that the radiative heating rate (\ref{eq:qradmonj}) can be expressed as
\begin{equation}
Q_\mathrm{rad} = 4 \pi \! \int_0^\infty \!  \kappa_\lambda \, (J_\lambda-B_{\!\lambda}) \, d\lambda.
\label{eq:qradmonjb}
\end{equation}
Equation~(\ref{eq:qradmonjb}) emphasizes that for radiative transfer with coherent scattering only absorption processes effectively contribute to radiative heat exchange. The consequences of this will be discussed in more detail below, in Sect.~\ref{sec:discuss}.

\begin{figure*}
\centering
\resizebox{\hsize}{!}{  \includegraphics{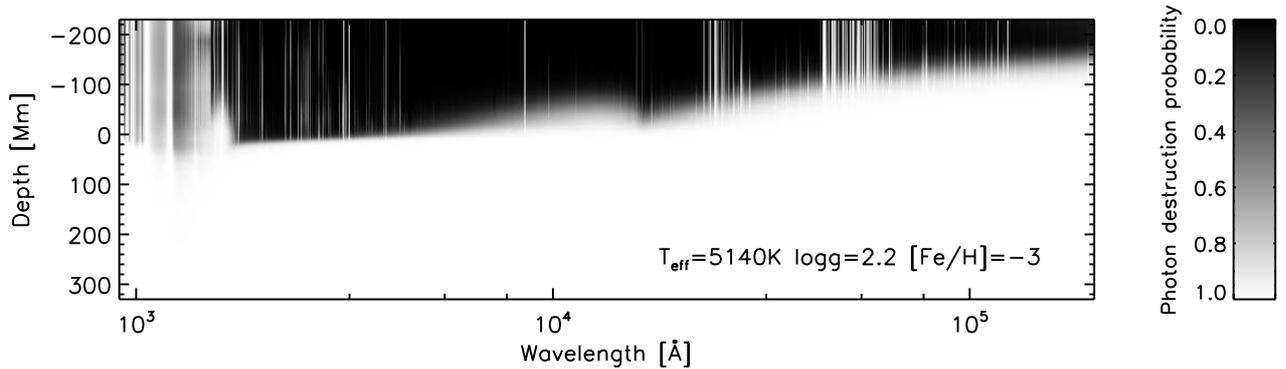}  }
\caption{Photon destruction probability as a function of wavelength and geometrical depth in the mean stratification of a metal-poor ([Fe/H]=$-3$) red giant surface convection simulation. Darker shades indicate lower photon destruction probability. The destruction probabilities are here computed assuming that only  continuous processes contribute to scattering extinction. Spectral lines (appearing as white vertical stripes in the upper part of the figure) are treated in pure absorption.}
\label{fig:epsilon1}
\end{figure*}
Continuum scattering plays a negligible role in shaping the photospheric stratification of solar-type stars \citep{hayek10}. In the much thinner atmospheres of metal-poor giants, however, where the electron number density is also lower, the absorption opacities as well as the rates of collisional processes that can thermalize the radiation field are strongly reduced and scattering becomes an important opacity source.
As an example, Fig.~\ref{fig:epsilon1} illustrates the dependence of the destruction probability $\epsilon_\lambda$ on wavelength and depth in the mean stratification taken from a metal-poor red giant surface convection simulation by \citet{collet07}.
Deep below the optical surface (located at $\approx 0$~Mm in the figure), absorption processes dominate the interaction between gas and photons, and the radiation field is fully thermalized (${\epsilon_\lambda}{=}{1}$). Above it, continuum scattering -- in particular,  Rayleigh scattering by H~{\sc i } in the UV and blue part of the visible spectrum and electron scattering at longer wavelengths -- becomes increasingly important as an opacity source, and the destruction probability $\epsilon_\lambda$ is closer to zero for a large part of the spectrum.

\subsection{Opacity binning}
\label{sec:opacbin}
Solving the radiative transfer equation for a large number of wavelengths in 3D convection simulations is a computationally demanding task.
We therefore use here the \emph{opacity binning} or \emph{multigroup} method \citep{nordlund82} 
%% {nordlund82,nordlund90,skartlien00} 
to approximate the dependence of opacities on wavelength. 
The problem of resolving the more than $100\,000$~wavelengths of the original opacity sampling data is reduced to computing the radiative transfer for only a small number of mean opacities.
Monochromatic continuous plus line opacities are sorted into four representative bins (groups) according to their strength. Bin membership is determined by the height where monochromatic optical depth unity is reached on the simulations' mean temperature-density stratifications. 
The latter are computed beforehand from thermally relaxed simulation sequences.\footnote{More specifically, we compute here the mean stratifications by averaging the simulation snapshots over time and over surfaces of constant column mass density. }
As the reference height scale we chose here the Rosseland optical depth computed for each simulation's mean stratification. The bin boundaries for the three simulations considered here are shown in Table~\ref{tab:bins}.

\begin{table}
\begin{minipage}[t]{\columnwidth}
\caption{Parameters of the 3D hydrodynamic simulations.}
\label{tab:param}
\centering
\renewcommand{\footnoterule}{}  % to avoid a line before footnotes
\begin{tabular}{ccccccc}
\hline\hline 
\noalign{\smallskip}
$\langle T_\mathrm{eff}\rangle$~\footnote{Temporal average of the emergent effective temperatures.}   &
$\log{g}$   & 
$[\mathrm{Fe/H}]$   & 
$x$,$y$,$z$-dimensions &
Time~\footnote{Time span of the simulations.} & 
$t_\mathrm{gran}$~\footnote{Average granule lifetime.} &
$t_\mathrm{rad}$~\footnote{Radiative heating timescale when switching to scattering as true absorption in the upper photosphere.} \\
$\mathrm{[K]}$	&
$[\mathrm{cgs}]$ 	&
 			& 
$\mathrm{[Mm]}$	&
$[\mathrm{hrs}]$  &
$[\mathrm{hrs}]$  &
$[\mathrm{hrs}]$  \\ 
\noalign{\smallskip}
\hline
\noalign{\smallskip}
%bifrost code Teffs:
${5140}$	& $2.2$	& $-3.0$ & $1150{\times}1150{\times}430$   &  $28$  & $20$ & $3.5$\\
${5080}$	& $2.2$	& $-2.0$ & $1150{\times}1150{\times}430$   &  $28$  & $20$ & $1.5$\\
${6510}$	& $4.04$	& $-3.0$ & $21.4{\times}21.4{\times}8.35$    &  $1.3$  & $0.3$ & $0.15$ \\
\noalign{\smallskip}
\hline 
\end{tabular}
\end{minipage}
\end{table}
\begin{table}
\begin{minipage}[t]{\columnwidth}
\caption{Bin boundaries: a wavelength $\lambda$ is assigned to a certain opacity bin depending on the Rosseland optical depth at which the \emph{monochromatic} optical depth $\tau_\lambda$ is equal to one. The table shows the Rosseland optical depth ranges spanned by individual bins in the present study.}
\label{tab:bins}
\centering
\renewcommand{\footnoterule}{}  % to avoid a line before footnotes
\begin{tabular}{p{0.15\linewidth}ccc}
\hline\hline 
\noalign{\smallskip}
Bin    &    \multicolumn{3}{c}{$\log\tau_\mathrm{Ross}(\tau_\lambda=1)$~range} \\
\noalign{\smallskip}
\hline
\noalign{\smallskip}
$1$   &  $+\infty$&   \ldots  &  $-0.5$   \\
$2$   &  $-0.5$   &    \ldots  & $-1.5$  \\
$3$   &  $-1.5$   &    \ldots  &  $-2.5$   \\
$4$   &  $-2.5$   &    \ldots  & $-\infty$ \\
\noalign{\smallskip}
\hline 
\end{tabular}
\end{minipage}
\end{table}

We followed the formalism of \citet{skartlien00} to compute the mean opacities and integrated thermal emissivities for each bin.
In computing the mean opacities, we distinguished between diffusion and streaming regimes, depending on whether  the averages are computed in the optically thick or in the optically thin layers of the mean temperature-density stratifications, respectively.
In the diffusion limit, we define the extinction coefficient in bin $i$ as the Rosseland group mean
\begin{equation}
\chi_i^{R}= \int_{\Delta\lambda_i} \! \frac{dB_\lambda}{dT} \, d\lambda   \,\Big/  \int_{\Delta\lambda_i}  \frac{1}{\chi_\lambda} \, \frac{dB_\lambda}{dT} \, d\lambda,
\label{eq:binross}
\end{equation}
where $\Delta\lambda_i$ denotes the set of wavelength intervals assigned to bin $i$.
%%Note: the set is not necessarily simply connected ("continuous" in Skartlien (2000) calls it)
In the streaming limit on the other hand, we approximate the extinction coefficient in bin $i$ with an intensity-weighted mean
\begin{equation}
\chi_i^{J}= \int_{\Delta\lambda_i} \! \chi_\lambda J_\lambda^\mathrm{pp} \, d\lambda \,\Big/  \int_{\Delta\lambda_i} \! J_\lambda^\mathrm{pp} \, d\lambda,
\label{eq:binstream}
\end{equation}
where $J_\lambda^\mathrm{pp}$ is the monochromatic mean intensity computed for the mean (plane-parallel) atmospheric temperature-density stratification and averaged over the whole solid angle.
In general, we define the mean extinction coefficient as a height-dependent linear combination of the values in the diffusion and streaming limits by using an exponential bridging
\begin{equation}
\chi_i = e^{-f\tau_i^R}\,\chi_i^J +  (1-e^{-f\tau_i^R})\,\chi_i^R
\label{eq:bridging}
\end{equation}
where $\tau_i^R$ is the Rosseland group mean optical depth computed for the mean stratification and $f$ is a constant factor that we simply set equal to $30$ in the present study.
The exponential bridging allows for a smooth transition from streaming to diffusion regime, while the choice of $f\!=\!30$ ensures that the streaming regime component is adequately suppressed below the optical surface.\footnote{We also performed test calculations with opacity tables computed for a value of $f$=2 of the bridging parameter and found that the resulting stratifications from the simulations are essentially unaltered.}
Bin opacities for arbitrary temperature and density pairs are then simply determined by constant extrapolation of $J^{\mathrm{pp}}$ on surfaces of constant Rosseland optical depth of the opacities calculated for the mean stratification.

In order to investigate the effective role of scattering on the radiative heating balance in 3D surface convection simulations, we considered three methods to handle it in the radiative transfer calculations.

As a default, we solve the radiative transfer equation consistently for the general case of a source function with a coherent isotropic continuum scattering term. We will refer to this as the standard case.
More precisely, for each bin we assume a source function of the form
\begin{equation}
S_i= ({\epsilon}B)_i + (1-\epsilon_i)J_i,
\label{eq:sourcef0}
\end{equation}
where $J_i$ is the mean intensity for bin $i$,
\begin{equation}
({\epsilon}B)_i=
\frac{1}{\kappa_i^{J}+\sigma_i^{J}}
\int_{\Delta\lambda_i} \! \kappa_\lambda B_\lambda \,d\lambda 
\label{eq:emiss0}
\end{equation}
is the integrated thermal emissivity,
\begin{equation}
\kappa_i^J=  {\int_{\Delta\lambda_i} \! \kappa_\lambda J_\lambda^\mathrm{pp} \,d\lambda } \,\Big/ { \int_{\Delta\lambda_i} \!J_\lambda^\mathrm{pp} \,d\lambda }
\label{eq:kappa0}
\end{equation}
and
\begin{equation}
\sigma_i^J=  {\int_{\Delta\lambda_i} \! \sigma_\lambda J_\lambda^\mathrm{pp} \,d\lambda } \,\Big/ { \int_{\Delta\lambda_i} \!J_\lambda^\mathrm{pp} \,d\lambda }
\label{eq:sigma0}
\end{equation}
are the streaming-regime mean absorption and mean scattering extinction coefficients, respectively, and 
\begin{equation}
\epsilon_i= \frac{\kappa_i^J}{\kappa_i^{J}+\sigma_i^{J}}
\label{eq:epsilon0}
\end{equation}
is the mean photon-destruction probability.
The mean intensity $J_i$ is determined by solving the radiative transfer equation for intensity at each time-step with the iterative method implemented by \citet{hayek10} and averaging over solid angle:
\begin{equation}
J_i = \frac{1}{4\pi}\,\sum_j\,\omega_j\,I_{ij},
\label{eq:meanint0}
\end{equation}
where $I_{ij}$ is the intensity in bin $i$ and along direction $j$, and the $\omega_j$'s are the weights of the Carlson's quadrature.
Finally, with the opacity binning approximation the radiative heating rate (\ref{eq:qradmonj})  becomes
\begin{equation}
Q_\mathrm{rad} =  \sum_i\!\sum_j\,\chi_i\,\omega_j\,( I_{ij} -S_i ) = 4\pi\!\sum_i\,\chi_i\,( J_i -S_i ).
\label{eq:qrad0}
\end{equation}

As a second approach, we approximated the monochromatic source function with the Planck function ($S_{\!\lambda}=B_\lambda$) throughout the simulation domain and completely excluded the contribution of continuum scattering from the calculation of the mean extinction coefficients in the streaming regime in each bin.
This is essentially the same approximation of scattering as the one adopted by \citet{nordlund90} \citep[see also][Sect.~5.2.1]{skartlien00} and subsequently in stellar surface convection simulations carried out by our group \citep[e.g.,][]{asplund99,asplund01,collet07} with the \citet{stein98} code, or by \citet{collet09} with the {\sc Stagger-Code} \citep{nordlund95}.
Compared with the standard case, this method is faster and computationally less demanding.

With this approximation, the integrated source function in bin $i$ is given by
\begin{equation}
S_i \,=\, B_i  \,= \int_{\Delta\lambda_i} \! B_\lambda \,d\lambda 
\label{eq:sourcef1}
\end{equation}
and the mean extinction coefficient in the streaming regime by
\begin{equation}
\chi_{i}^{J\,\ast}=\kappa_i^J=  {\int_{\Delta\lambda_i} \! \kappa_\lambda J_\lambda^\mathrm{pp} \,d\lambda } \,\Big/ { \int_{\Delta\lambda_i} \!J_\lambda^\mathrm{pp} \,d\lambda }.
\label{eq:chiJ1}
\end{equation}
In the opacity binning approximation, the radiative heating rate (\ref{eq:qradmon}) becomes
\begin{equation}
Q_\mathrm{rad}^{\,\ast} = Q_\mathrm{rad} =4 \pi \!  \sum_i \,  \chi_i^{\ast} \, ( J_i -B_i ),
\label{eq:qrad1}
\end{equation}
where $J_i$ is again the mean intensity for bin $i$.
With our assumption (\ref{eq:chiJ1}) for the extinction coefficient in the streaming regime, 
the radiative heating rate in the optically thin layers of the convection simulations approaches
\begin{equation}
Q_\mathrm{rad}^{\,\ast} \, {\approx}\, 4 \pi \!  \sum_i \,  \kappa_i^J \, ( J_i -B_i ).
\label{eq:qrad1stream}
\end{equation}

Finally, as a third approach, we still assumed a Planckian source function, but this time included the contribution of scattering to the total extinction in both the diffusion and the streaming regimes, effectively treating scattering as true absorption everywhere in the simulation domain. 
This corresponds to the approximation adopted in stellar surface convection simulations by the {\sc CO$^5$BOLD} collaboration (H.~G.~Ludwig, priv. comm.).

For each set of stellar parameters indicated in Table~\ref{tab:param}, starting from the same initial snapshot, we computed three simulation sequences -- one for each of the different treatments of the radiative transfer considered above -- and followed the evolution of the temperature stratification with time in all three cases.

\begin{figure*}
\centering
\resizebox{\hsize}{!}{
  \includegraphics{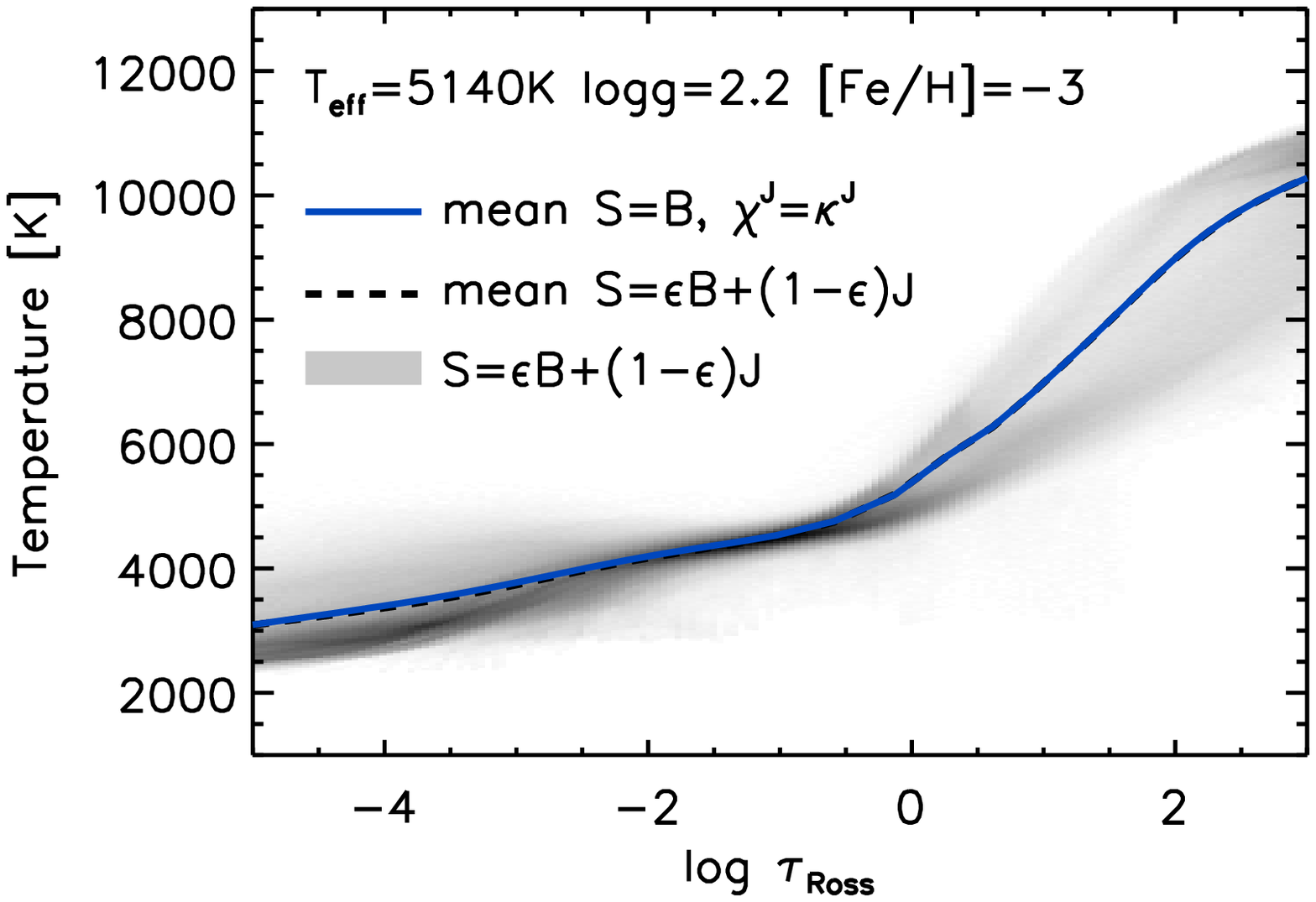}
  \includegraphics{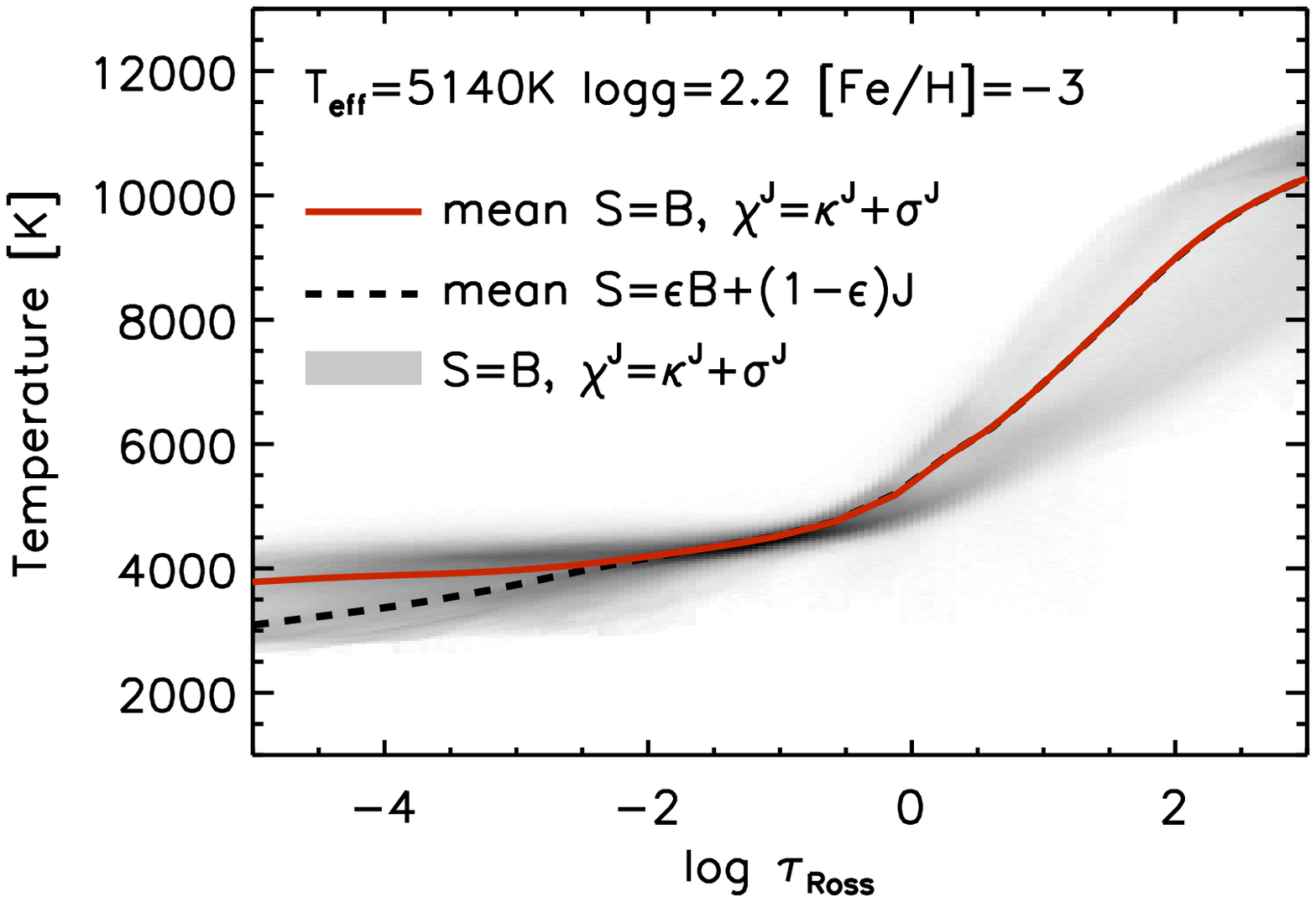} }
  \caption{\emph{Grey shaded areas}: distribution of temperature as a function of Rosseland optical depth in two representative snapshots of two surface convection simulations of a metal-poor red giant at [Fe/H]=$-3$. The snapshots were taken $28$~hours after the start of the simulations. Darker shades indicate temperature values with higher frequency of occurrence. 
 \emph{Left panel}: temperature distribution from the simulation with self-consistent solution of the radiative transfer equation for a source function with a coherent scattering term (standard case).
 \emph{Continuous blue line}: mean temperature stratification from the corresponding simulation assuming 
 a Planckian source function and neglecting the contribution of scattering in optically thin layers, averaged over time and surfaces of constant Rosseland optical depth. 
 \emph{Right panel}: temperature distribution for the radiative transfer solution assuming a Planckian source function and treating scattering as true absorption everywhere.
  \emph{Continuous red line}: corresponding mean temperature stratification.
 \emph{Dashed lines (both panels)}: mean temperature stratification for the standard-case simulation. }
  \label{fig:atmos1}
\centering
\resizebox{\hsize}{!}{
	\includegraphics{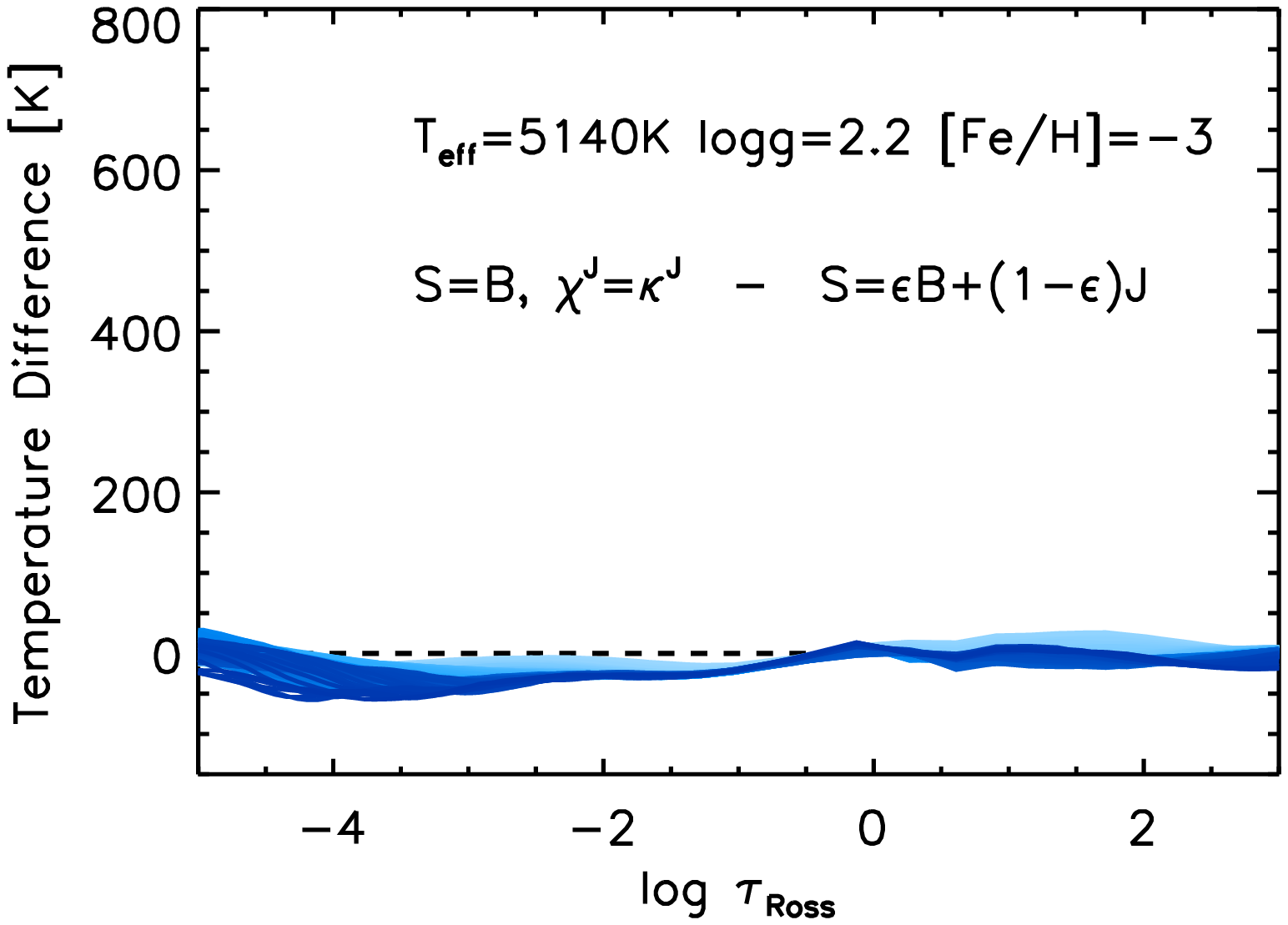} 
	\includegraphics{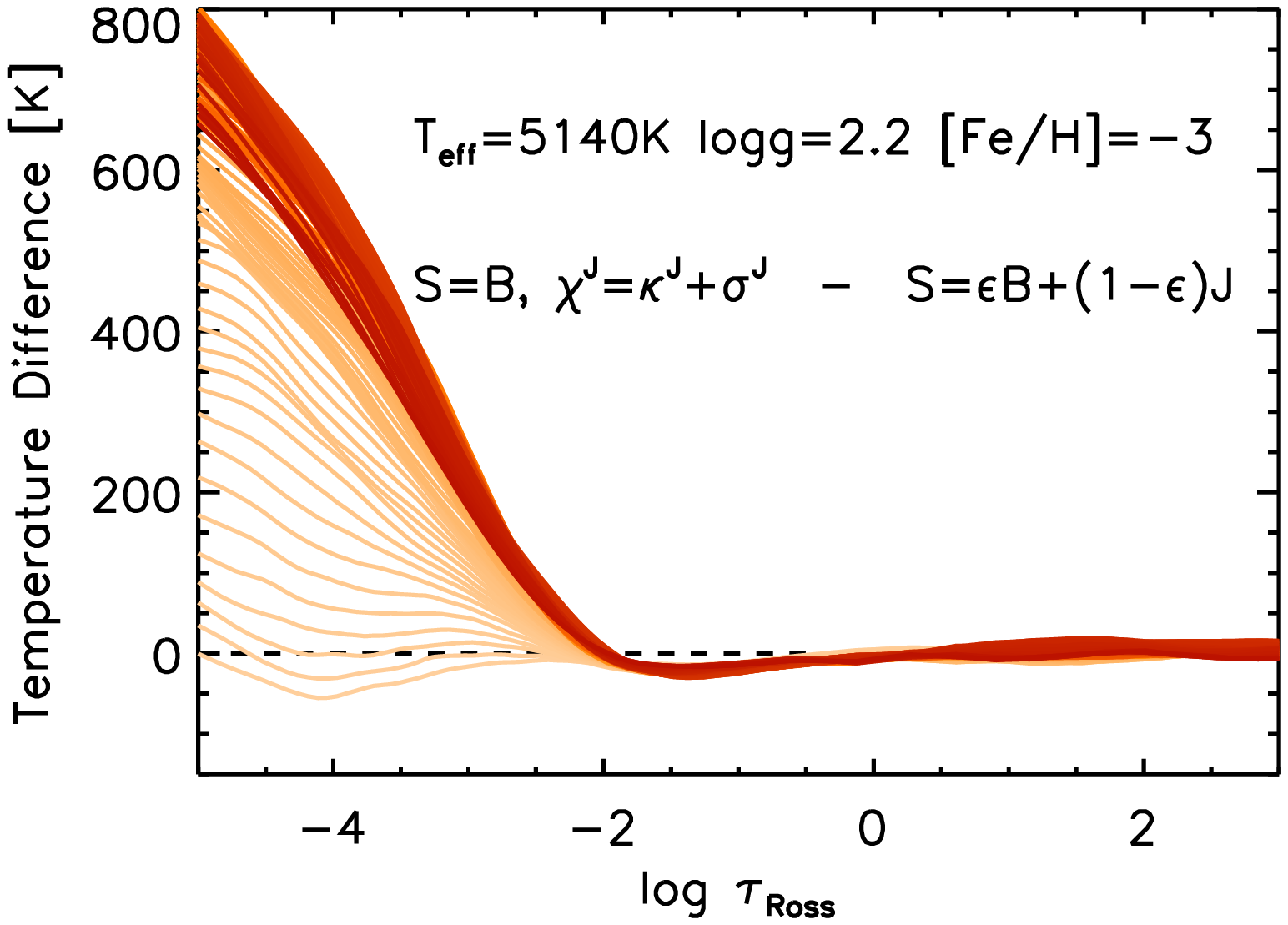}
}
\caption{Temporal evolution of the mean temperature stratification from the two metal-poor red giant ([Fe/H]=$-3$) surface convection simulations considered in Fig.~\ref{fig:atmos1}. 
The curves show the \emph{differences} of the mean temperature with respect to the standard case (self-consistent solution of the radiative transfer equation for a source function with a coherent scattering term) as a function of time. The instantaneous mean temperature stratifications are computed on surfaces of constant Rosseland optical depth.
The total temporal coverage is $28$~hours, with snapshots taken every 10~minutes. Darker curves indicate later snapshots in the two sequences.
\emph{Left panel}:  temporal evolution of the temperature differences for the simulation computed assuming a Planckian source function and neglecting the contribution of scattering to total extinction in optically thin layers. \emph{Right panel}: corresponding evolution for the simulation computed under the assumption of Planckian source function with scattering treated as true absorption everywhere.}
\label{fig:tmean1}
\end{figure*}

\begin{figure*}
\centering
\resizebox{\hsize}{!}{ 
	\includegraphics{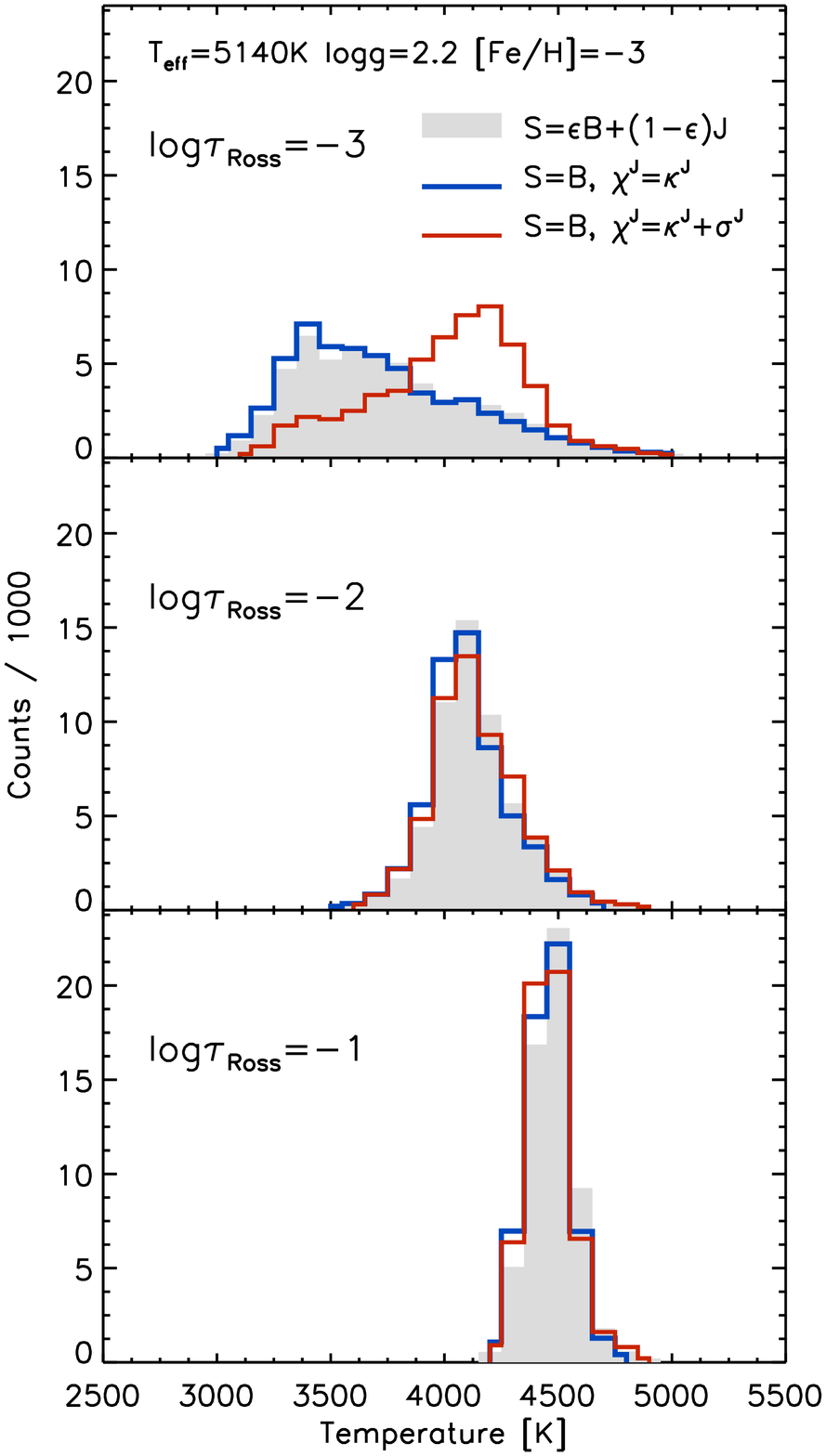} 
	\includegraphics{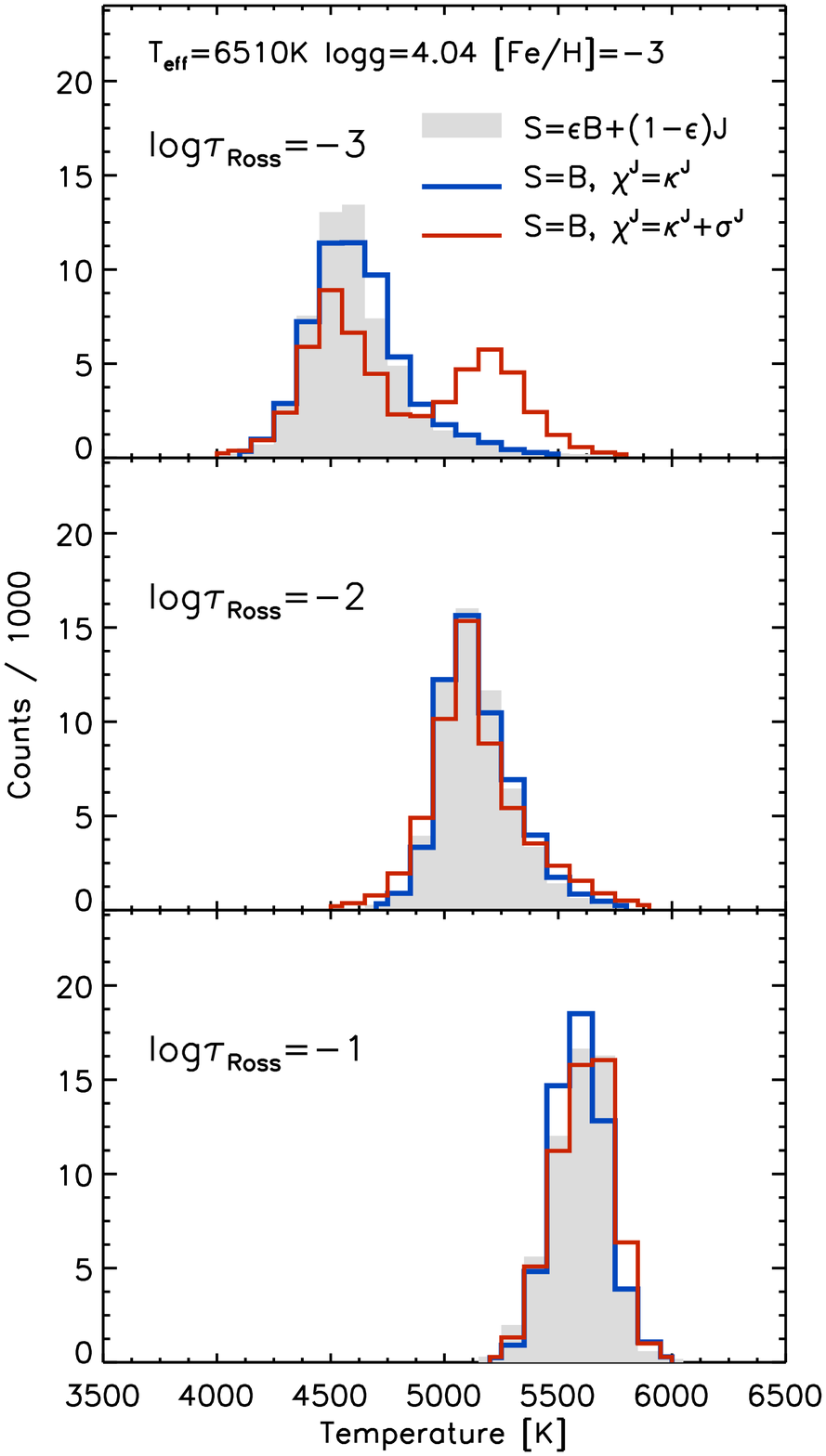}
	\includegraphics{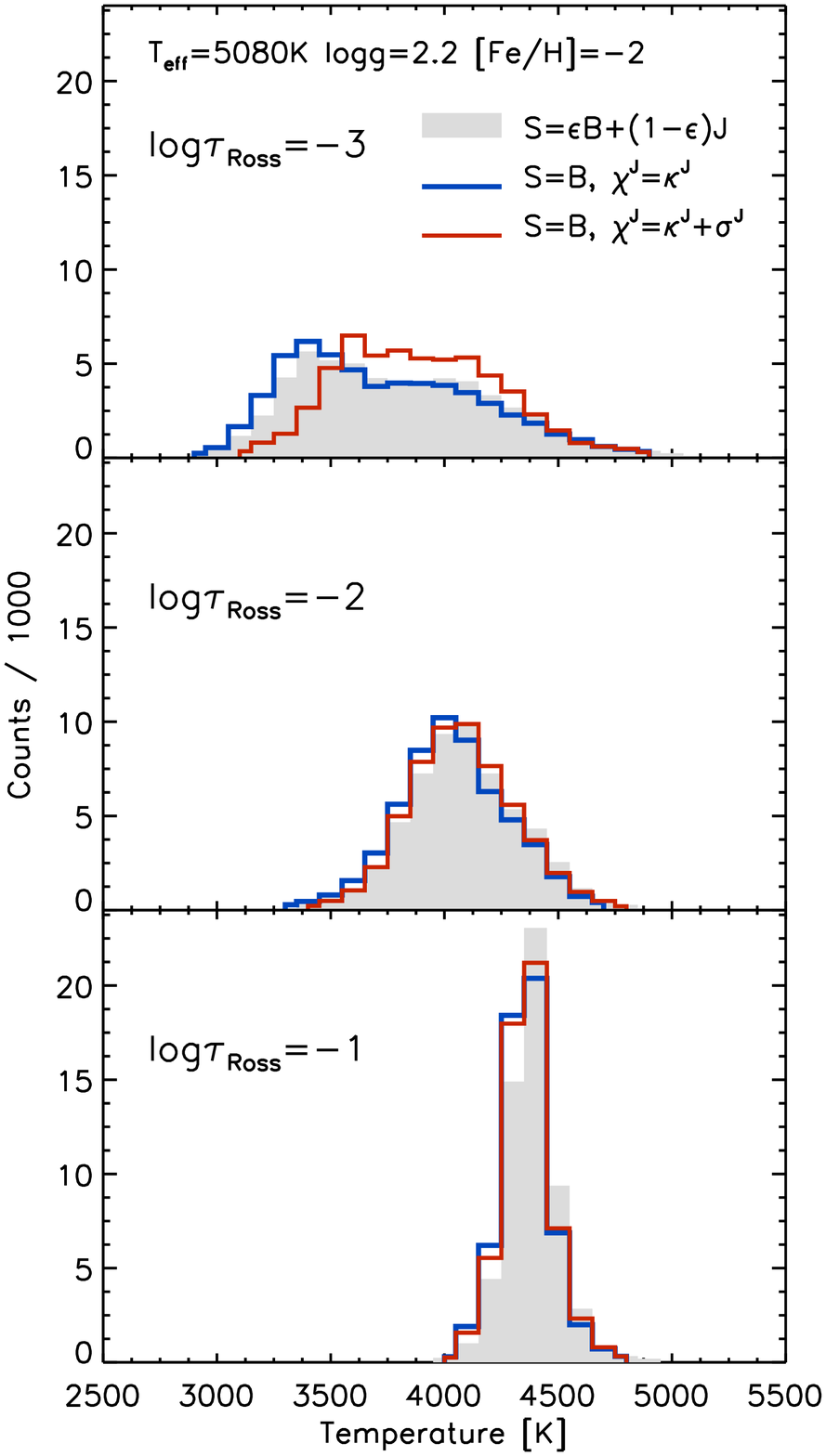} 
	}
\caption{Histograms of temperature in the upper photospheric layers of the surface convection simulations at three different Rosseland optical depths and for the three different treatments of scattering in the radiative transfer solution. 
\emph{Left panel}: results for the red giant at [Fe/H]=$-3$; \emph{central panel}: turnoff star at [Fe/H]=$-3$; \emph{right panel}: red giant at [Fe/H]=$-2$.
The histograms are constructed from snapshots taken at the end of the simulation sequences, for the coherent scattering case (\emph{grey bars}), Planckian source function and no scattering extinction in optically thin layers (\emph{blue curves}), and Planckian source function with scattering treated as true absorption (\emph{red curves}).}
\label{fig:temphist123}
\end{figure*}

%__________________________________________________________________
\section{Results}

Figure~\ref{fig:atmos1} compares the resulting temperature structures from the [Fe/H]=$-3$ red giant surface convection simulations computed for the three cases: (i) self-consistent solution of the radiative transfer solution for a source function with a coherent scattering term (standard case); (ii) Planckian source function and no scattering contribution to extinction in optically thin layers; (iii) scattering treated as true absorption everywhere.
The temperature stratification resulting from the assumption of a Planckian source function and of no contribution of scattering to extinction in optically thin layers is very similar to the one predicted in the standard case (Fig.~\ref{fig:atmos1}, left panel).
In particular, the overall temperature distribution with optical depth is consistent with the previous results by \citet{collet07}.
As shown in the left panel of Fig.~\ref{fig:tmean1}, the snapshot-by-snapshot temperature difference between the mean stratifications in the standard case and in the no-scattering-in-streaming-regime case is typically less than $50$~K at all depths.
Moreover, not only the mean stratifications resulting from these two treatments of scattering in the solution of the radiative transfer equation are virtually identical, but the two simulations also maintain a good correlation in terms of 3D temperature and density structures as well as velocity fields after several ($\approx$~12) hours of simulated stellar time (the average granule lifetime is about $20$~hours).

Conversely, including scattering as true absorption while assuming a Planckian source function causes a rapid heating of the upper photospheric layers immediately after starting the simulation (Fig.~\ref{fig:tmean1}, right panel).
With respect to the standard simulation, the temperature at $\log\tau_\mathrm{Ross}$=$-4$ rises by $500$~K in about six hours of simulated stellar time.  As the simulation evolves, the absolute temperature difference with respect to the standard case settles to about $600$~K at that depth (Fig.~\ref{fig:atmos1}, right panel). The temperature rise with respect to the standard case is approximately proportional to $1-\exp(-t/t_\mathrm{rad})$, where $t_\mathrm{rad}$ is the characteristic radiative heating timescale of the upper photospheric layers when scattering is switched to true absorption.
In particular, for the [Fe/H]=$-3$ red giant simulation, $t_\mathrm{rad}{\approx}3~$hours; the characteristic radiative heating timescales for the other simulations are listed in Table~\ref{tab:param}.
Mixing caused by bulk gas flows has an effect on the temperature adjustments in the atmospheres, but on longer timescales.
At greater depths the differences are less pronounced and amount to about $300$~K at $\log\tau_\mathrm{Ross}$=$-3$ when the temperature balance is achieved and to less than $50$~K below $\log\tau_\mathrm{Ross}$=$-2$. This dependence of the heating on depth also results in a shallowing of the vertical temperature gradient in the upper photosphere. 
Near the optical surface, the temperature differences are negligible and the effective temperature of the simulation is practically unaltered.

Figure~\ref{fig:temphist123}, left panel, shows the temperature distribution at three different optical depths for the relaxed  [Fe/H]$=-3$ red giant simulations computed with the three different treatments of scattering. Again, the distribution returned by the simulation with proper treatment of coherent scattering agrees very well at all optical depths with the one predicted by the standard case, but differs significantly from the one predicted in the scattering-as-absorption case. In particular, at an optical depth of $\log\tau_\mathrm{Ross}$=$-3$, where the differences between the scattering-as-absorption and coherent scattering cases are more evident, the latter predicts a skewed temperature distribution peaking at $3400$~K, with a longer tail at higher temperatures and with the hint of a smaller secondary peak at $4000$~K; the simulation that assumes scattering in absorption instead yields a distribution skewed towards higher temperatures and peaking at $4200$~K and with a less pronounced peak at around $3400$~K. The overall shape of the distribution is determined by the competition between dynamic and radiative heating/cooling processes. Radiative heating/cooling reduces the skewness of the distribution by attenuating temperature fluctuations. The secondary peak as well as the low-temperature tail of the distribution in these optically thin layers are mainly contributed by cool inflowing gas from the top of the simulation domain. At greater depths, the differences in temperature distribution between the three simulations become less and less significant.

\begin{figure*}
\centering
\resizebox{\hsize}{!}{
	\includegraphics{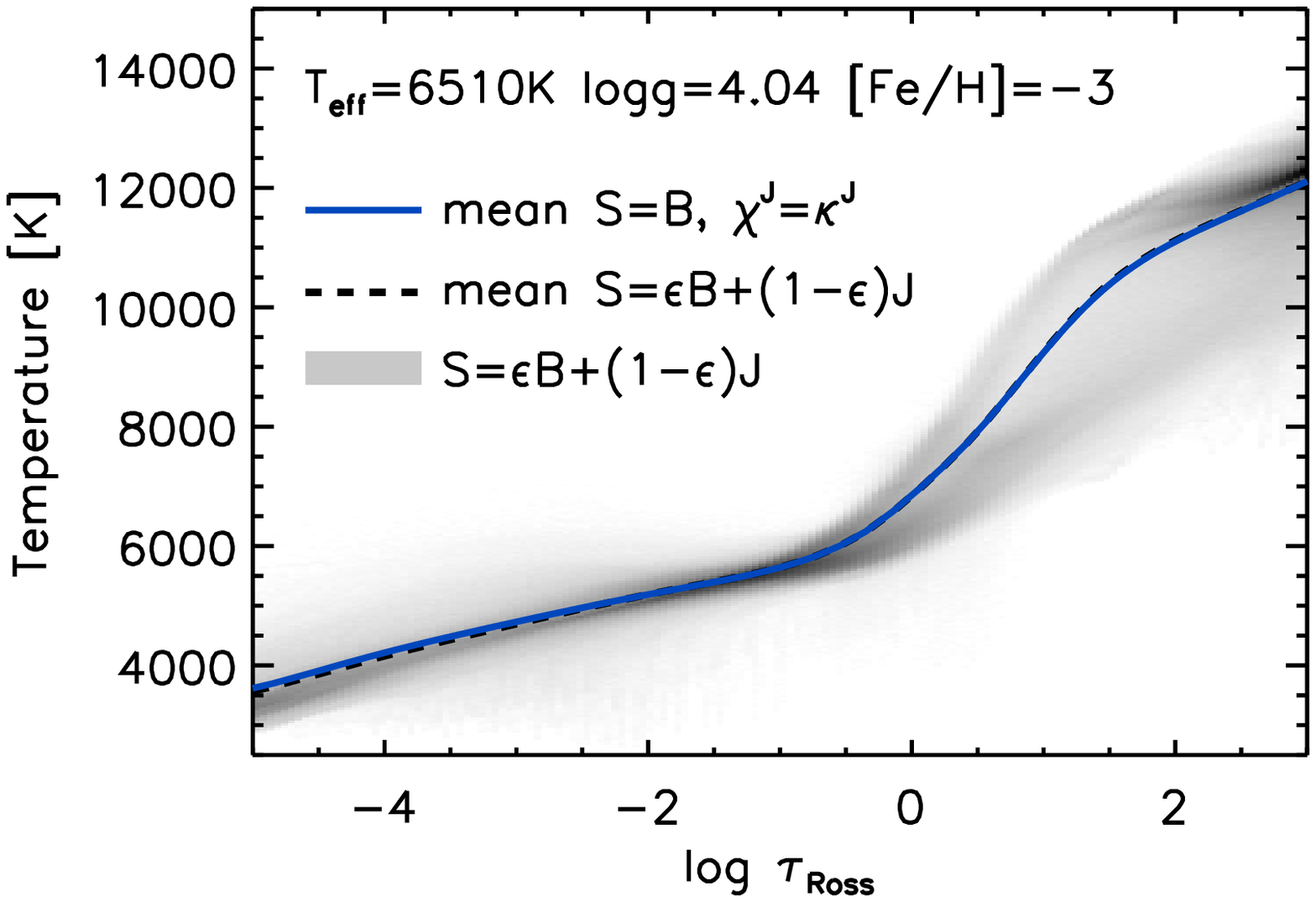}
	\includegraphics{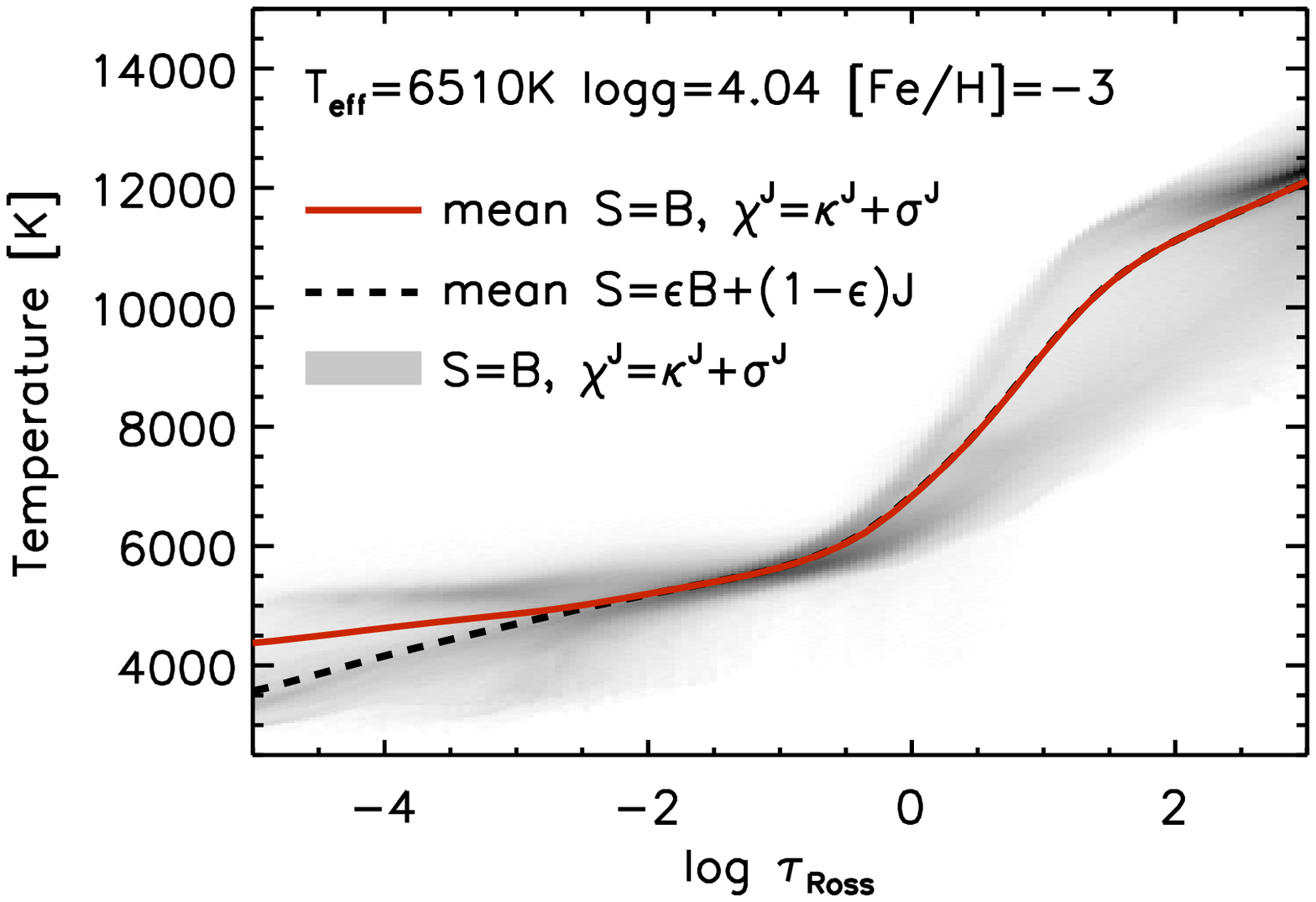} }
\caption{\emph{Left panel}: temperature distribution with Rosseland optical depth from a representative snapshot of the [Fe/H]=$-3$ turnoff-star surface convection simulation computed solving the radiative transfer equation self-consistently for a source function with a coherent scattering term (standard case).
Darker shades indicate higher probability. 
\emph{Blue curve}: mean stratification from the corresponding simulation computed assuming a Planckian source function and no contribution of scattering to total extinction in optically thin layers.
\emph{Right panel}: corresponding temperature distribution with Rosseland optical depth from the simulation assuming a Planckian source function and scattering as true absorption everywhere (mean temperature stratification shown in red).
\emph{Dashed lines (both panels)}: mean temperature stratification from the standard-case simulation.
All snapshots were taken $80$~minutes after the start of the simulations.}
\label{fig:atmos2}
\resizebox{\hsize}{!}{
	\includegraphics{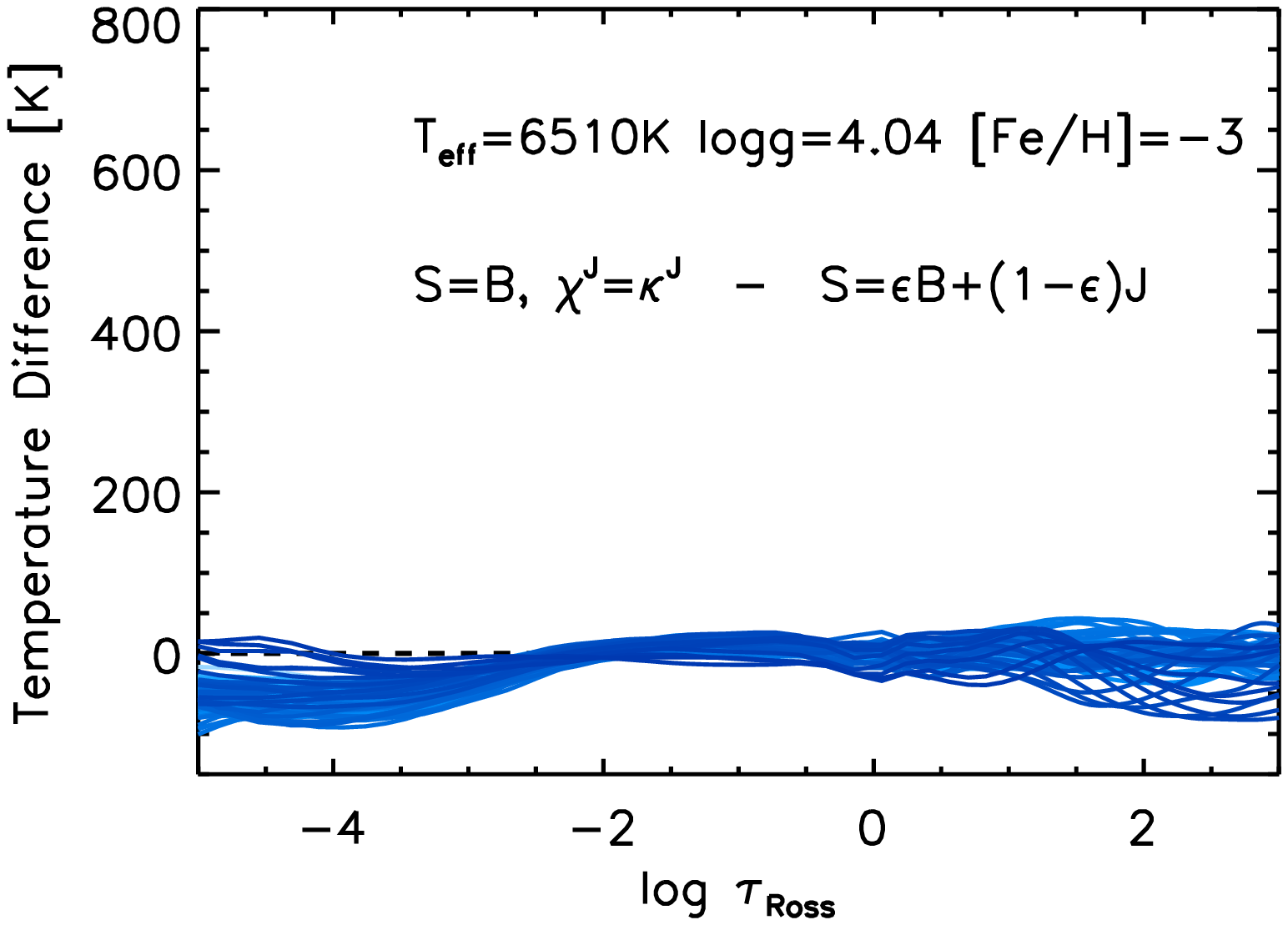} 
	\includegraphics{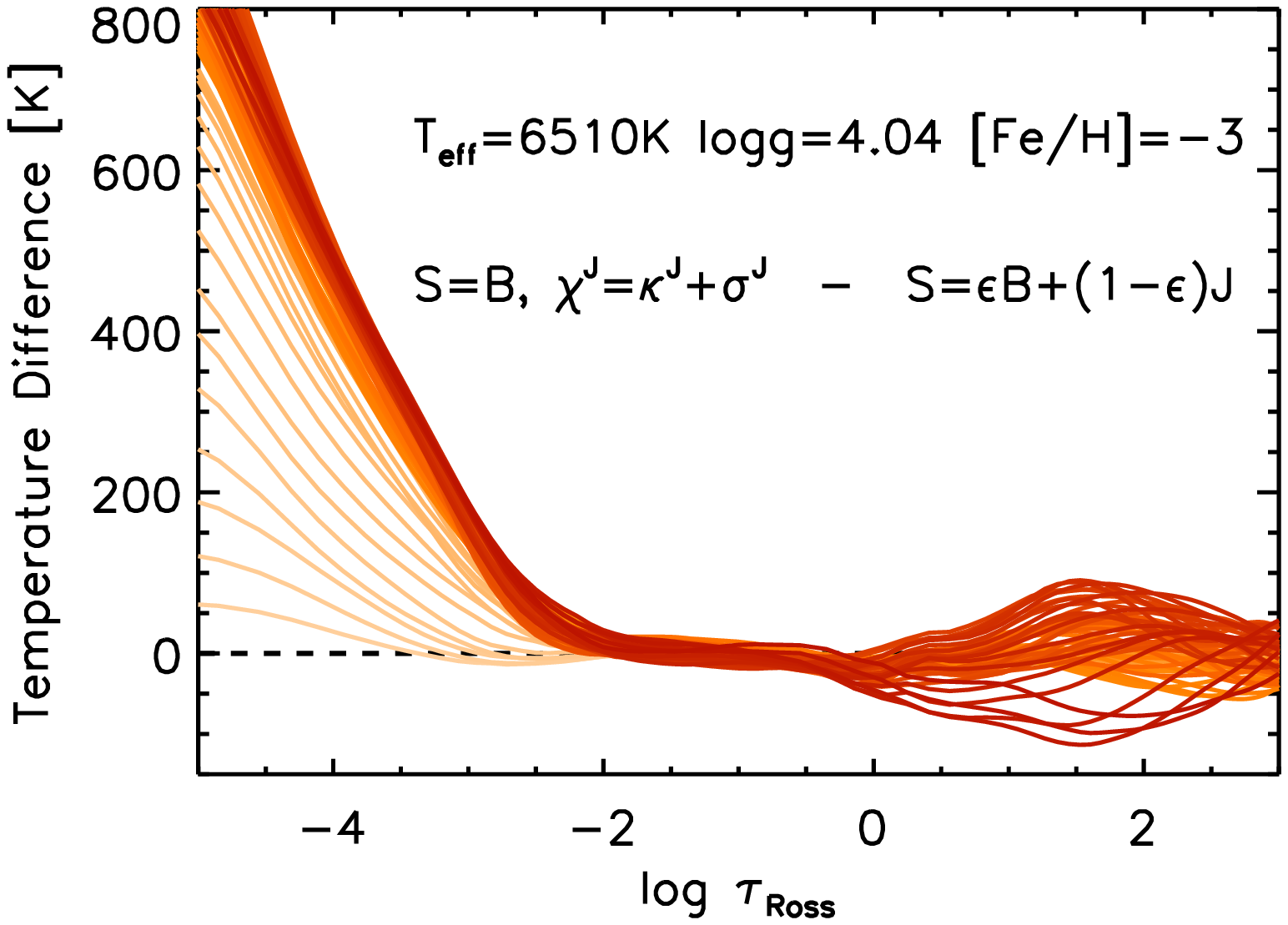}
}
\caption{Temporal evolution of the mean temperature stratification from the two metal-poor ([Fe/H]=$-3$) turnoff-star surface convection simulations considered in Fig.~\ref{fig:atmos2}. 
The curves show the \emph{differences} of the mean temperature with respect to the standard case (self-consistent solution of the radiative transfer equation for a source function with a coherent scattering term) as a function of time.
The total temporal coverage is $80$~minutes, with snapshots taken every minute. Darker curves indicate later snapshots in the two sequences.
\emph{Left panel}: temporal evolution of the temperature differences for the simulation assuming a Planckian source function and no contribution of scattering to total extinction in optically thin layers.
\emph{Right panel}: corresponding evolution under the assumption of Planckian source function with scattering treated as true absorption.}
\label{fig:tmean2}
\end{figure*}

\begin{figure*}
\centering
\resizebox{\hsize}{!}{
	\includegraphics{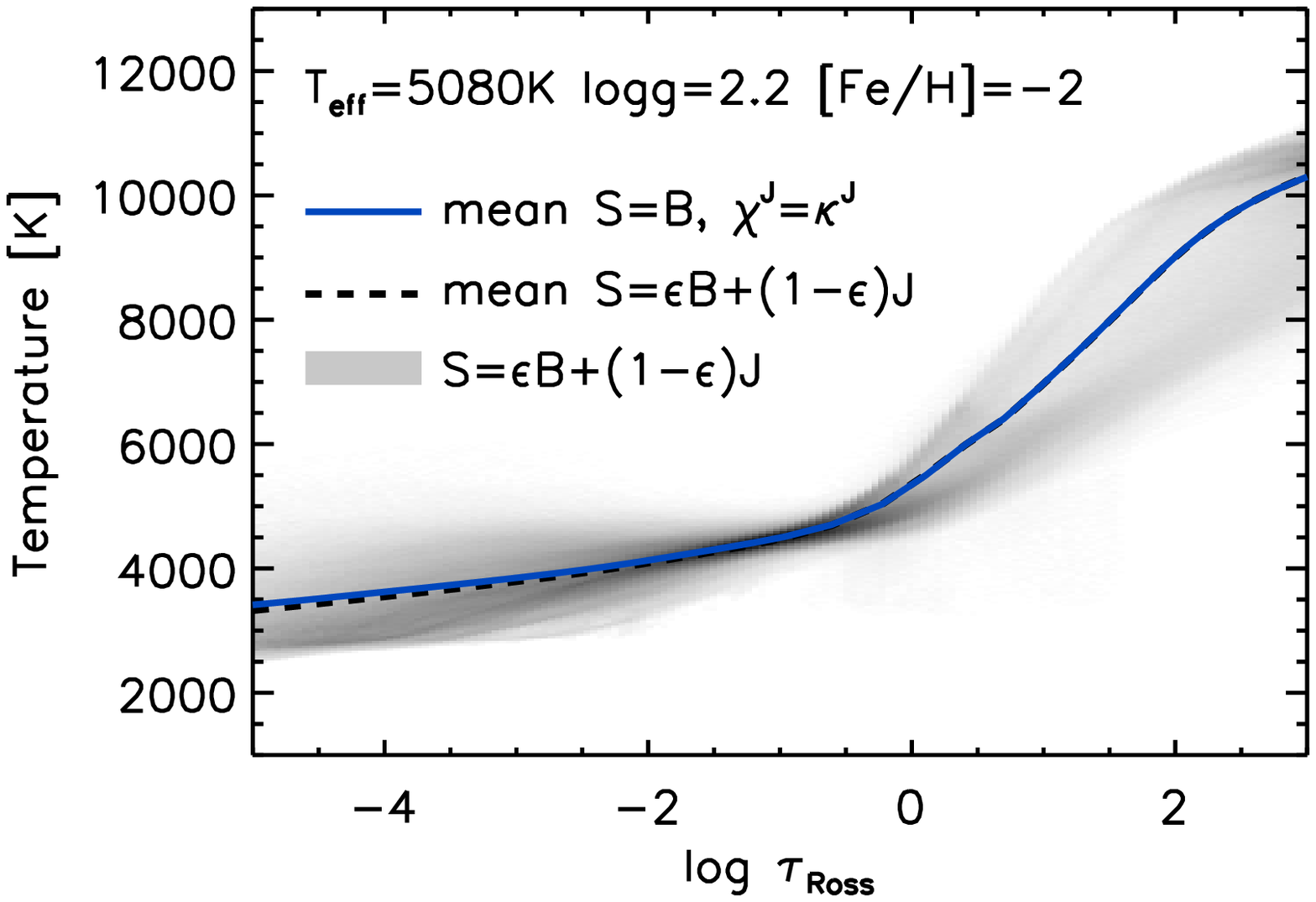}
	\includegraphics{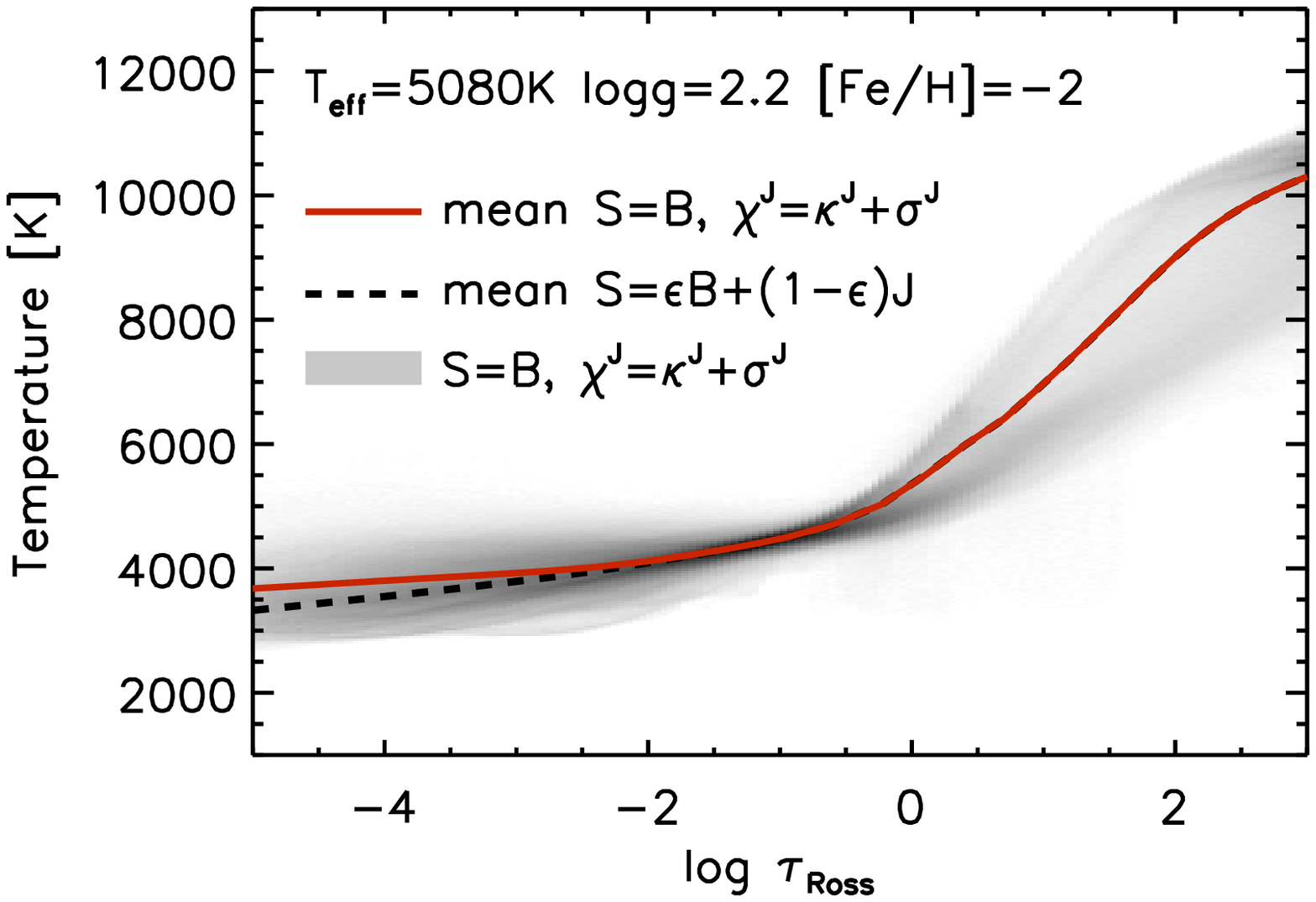}
}
\caption{Temperature distribution with Rosseland optical depth at the surface of a metal-poor red giant at [Fe/H]=$-2$. 
%%%
%%%
%%%
\emph{Left panel}: predicted distribution from a representative snapshot of the simulation computed solving the radiative transfer equation self-consistently for a source function with a coherent scattering term (standard case).
\emph{Blue curve}: mean stratification from the corresponding simulation computed assuming a Planckian source function and excluding scattering from extinction in optically thin layers.
\emph{Right panel}: corresponding distribution from a simulation assuming a Planckian source function with scattering as true absorption (mean stratification shown in red).
\emph{Dashed lines (both panels)}: mean stratification from the standard-case simulation. 
All snapshots were taken $28$~hours after the start of the simulations. See Fig.~\ref{fig:atmos1} for more details.}
\label{fig:atmos3}
\resizebox{\hsize}{!}{
	\includegraphics{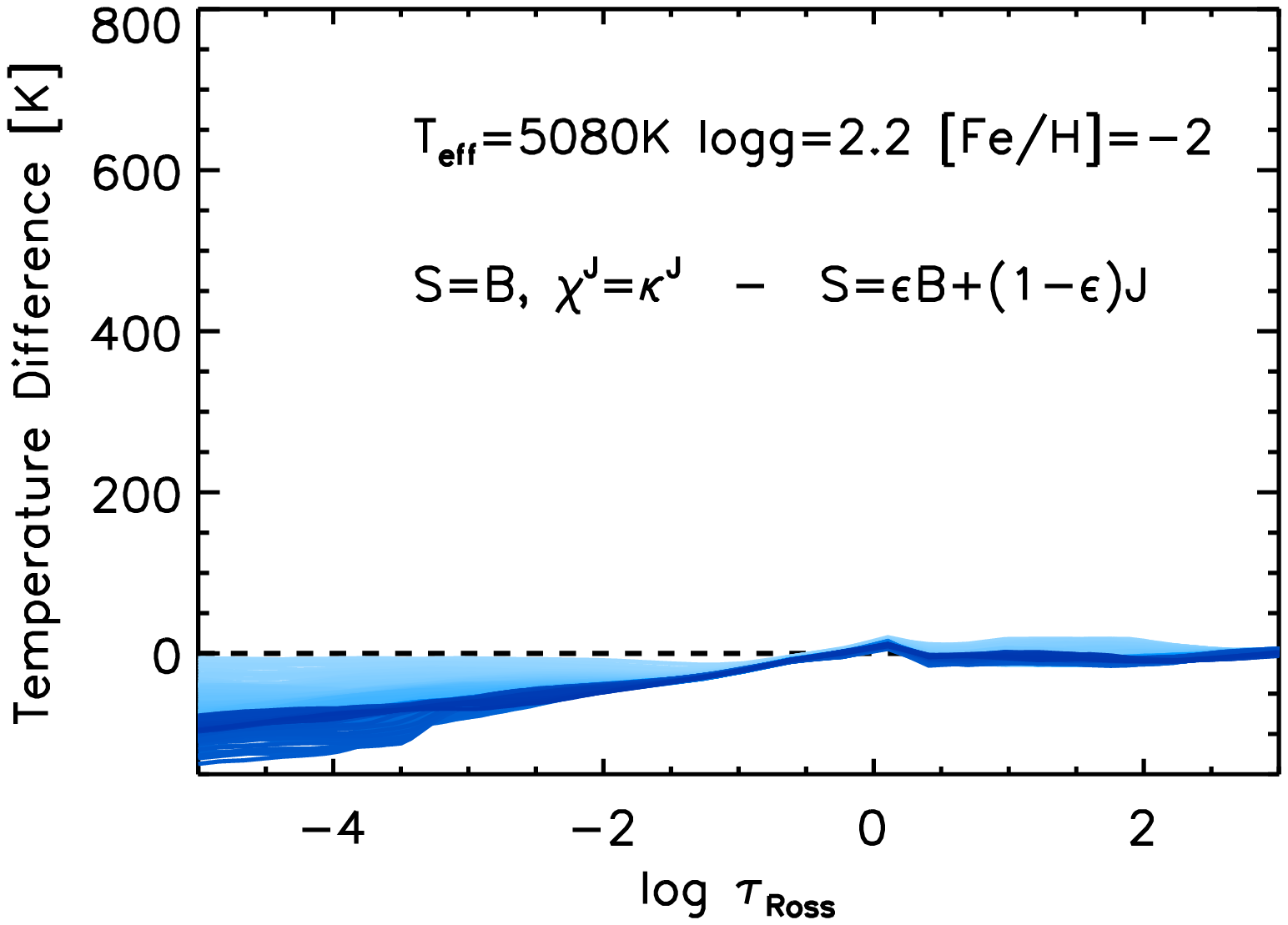} 
	\includegraphics{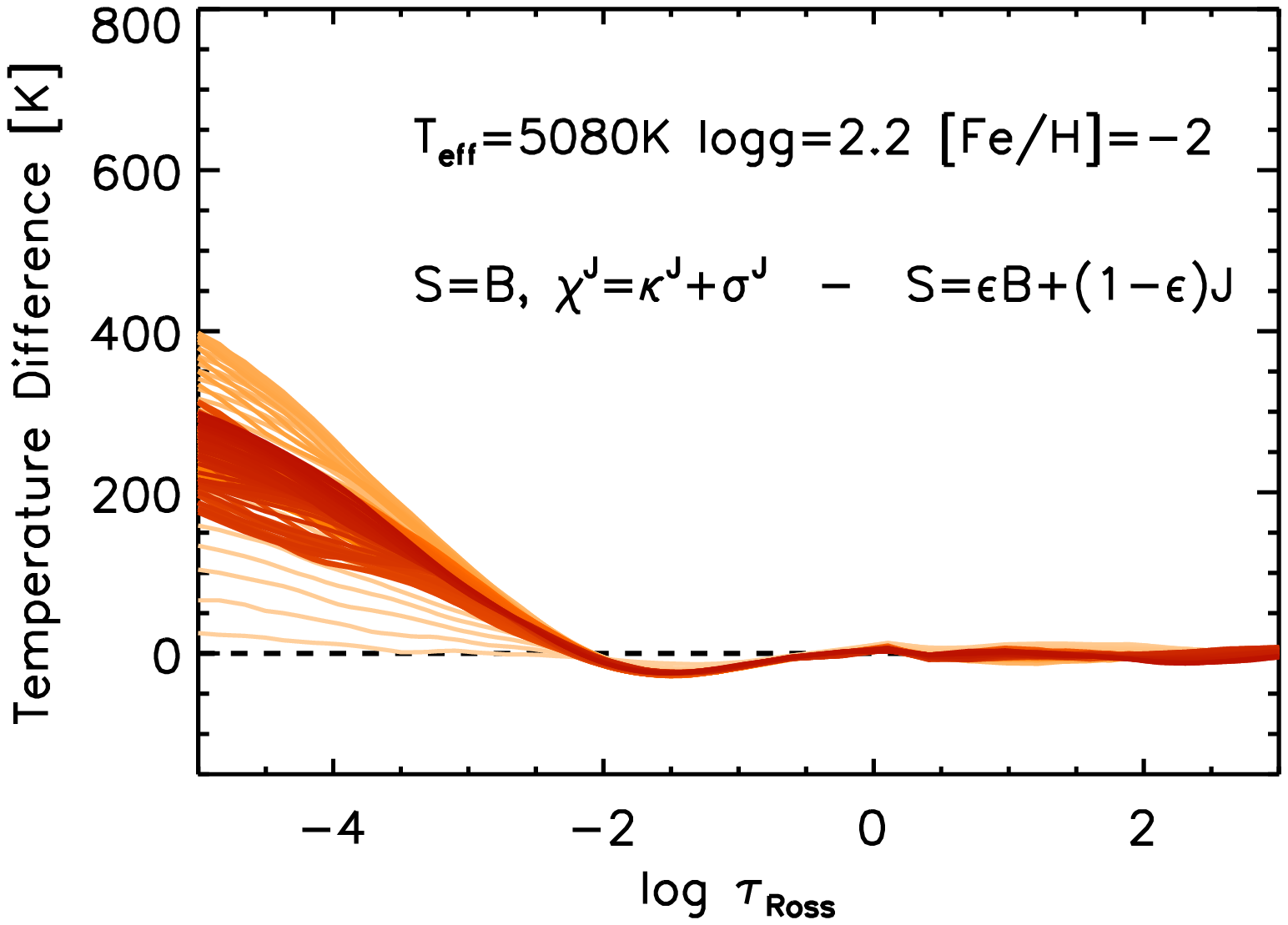}
}
\caption{Temporal evolution of the mean temperature stratification in the two metal-poor red giant simulations at [Fe/H]=$-2$ from Fig.~\ref{fig:atmos3}. 
The curves show the \emph{differences} of the mean temperature with respect to the standard case (self-consistent solution of the radiative transfer equation for a source function with a coherent scattering term) as a function of time.
The total temporal coverage is $28$~hours, with snapshots taken every hour. 
\emph{Left panel}: temporal evolution for the simulation assuming a Planckian source function and no contribution of scattering to the total extinction in optically thin layers.
\emph{Right panel}: corresponding evolution under the assumption of Planckian source function with scattering treated as true absorption.   See Fig.~\ref{fig:tmean1} for more details.}
\label{fig:tmean3}
\end{figure*}

Figure~\ref{fig:atmos2} shows the resulting distribution of temperature with optical depth for the metal-poor  ([Fe/H]=$-3$) turnoff-star simulations computed in the standard (left panel) and scattering-as-absorption (right panel) cases. As for the metal-poor red giant simulations at [Fe/H]=$-3$, the stratification predicted by the standard case agrees well with the one predicted by the simulation that assumes a Planckian source function and no scattering contribution to extinction in optically thin layers, while the case where scattering is treated as pure absorption everywhere produces a hotter and shallower temperature stratification in the upper photosphere. The corresponding histograms in the central panel of Fig.~\ref{fig:temphist123} illustrate in more detail the temperature distributions at various depths in the three cases. Again, the distributions in the standard and in the no-scattering-in-streaming-regime cases are very similar at all depths, while the temperature distribution for the simulation in which scattering is treated as pure absorption clearly departs from the other two in the uppermost photospheric layers.

The temporal evolution of the mean temperature stratifications is shown in Fig.~\ref{fig:tmean2}. Because of the stronger surface gravity and smaller spatial dimensions, the adjustment timescales are shorter for the turnoff-star simulations than for the red giant series ($t_\mathrm{rad}{\approx}10$~minutes for the turnoff sequence). The outer layers of the simulation in which scattering is treated as pure absorption heat up rapidly, and reach a stable configuration within approximately a few tens of minutes of stellar time, with the temperature being on average $600$~K and $200$~K higher than in the simulation with coherent scattering at $\log\tau_\mathrm{Ross}$=$-4$ and at $\log\tau_\mathrm{Ross}$=$-3$, respectively.
Because the dynamical adjustment timescales are short, the temperature and density perturbations arising from the different treatments of scattering in the radiative transfer grow rapidly and the correlation between simulations from the two different cases degrades relatively fast. This is partly the reason why the snapshot-by-snapshot temperature differences between the mean stratifications in the standard and scattering-in-absorption cases above the optical surface are slightly larger than for the [Fe/H]=$-3$ red giant simulations, although the differences typically remain below $100$~K.

We also carried out a series of simulations of a metal-poor red giant at slightly higher metallicity ([Fe/H]=$-2$) for the three different treatments of scattering. The results (Fig.~\ref{fig:atmos3} and~\ref{fig:tmean3}; Fig.~\ref{fig:temphist123}, right panel) are qualitatively similar to the [Fe/H]=$-3$ simulations. The scattering-in-absorption case also produces a hot stratification in the uppermost part of the photosphere, but the deviations from the standard case are smaller (${\approx}250$~K at $\log\tau_\mathrm{Ross}$=$-4$). Excluding scattering from the total extinction in the optically thin layers again leads to a stratification that agrees well with the coherent scattering case at all depths. However, in the upper part of the photosphere, the stratification from the simulation assuming a Planckian source function and no contribution of scattering to extinction in optically thin layers is on average slightly cooler (by $100$~K at $\log\tau_\mathrm{Ross}$=$-4$) than the one predicted in the standard case. 

%__________________________________________________________________
\section{Discussion}
\label{sec:discuss}

\begin{figure*}
\centering
\resizebox{\hsize}{!}{	\includegraphics{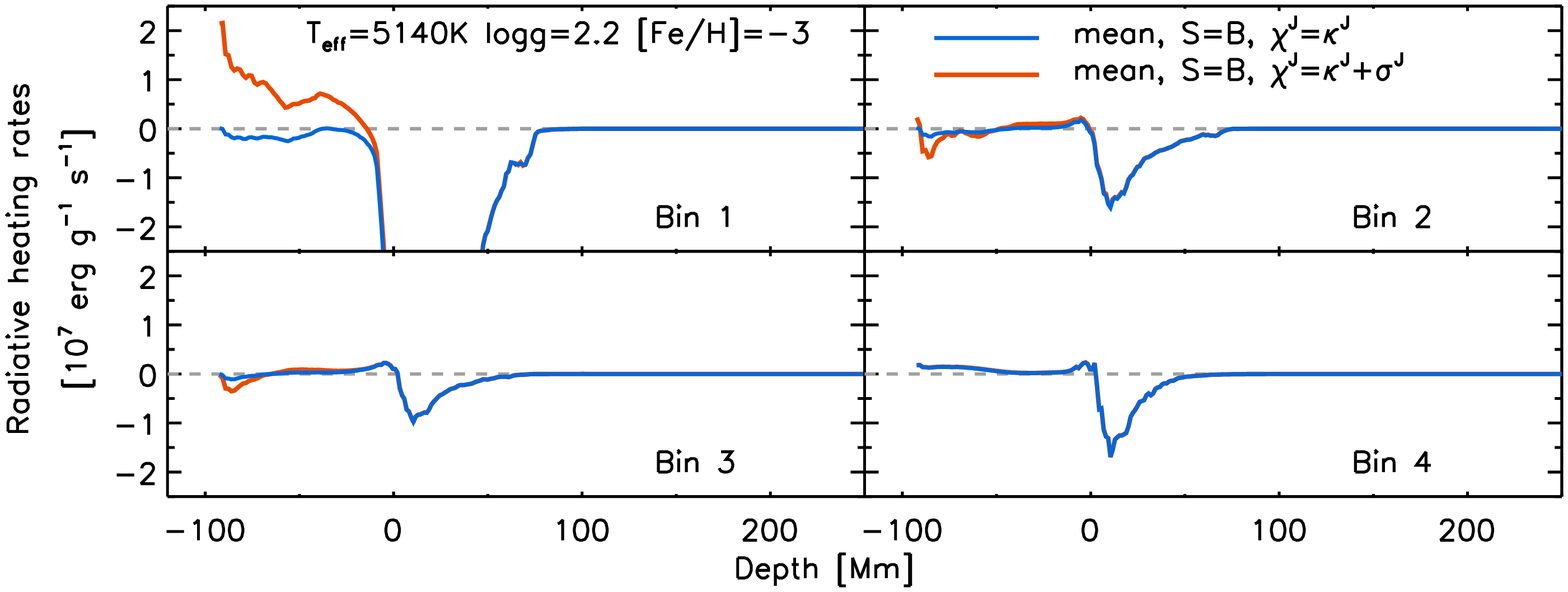} 	}
  \caption{\emph{Blue curves}: horizontally averaged radiative heating rates for individual bins as a function of geometrical depth in a snapshot of the [Fe/H]=$-3$ red giant simulation computed assuming a Planckian source function and excluding the contribution of scattering to total extinction in optically thin layers. For clarity, only the relevant (upper) part of the simulation is shown; radiative heating rates at the cooling peak are unchanged when switching between scattering treatments. \emph{Red curves}: horizontally averaged heating rates for the same snapshot immediately after switching to scattering as true absorption.}
\label{fig:qradmgiant}
\end{figure*}

\begin{figure*}
\centering
\resizebox{\hsize}{!}{
	\includegraphics{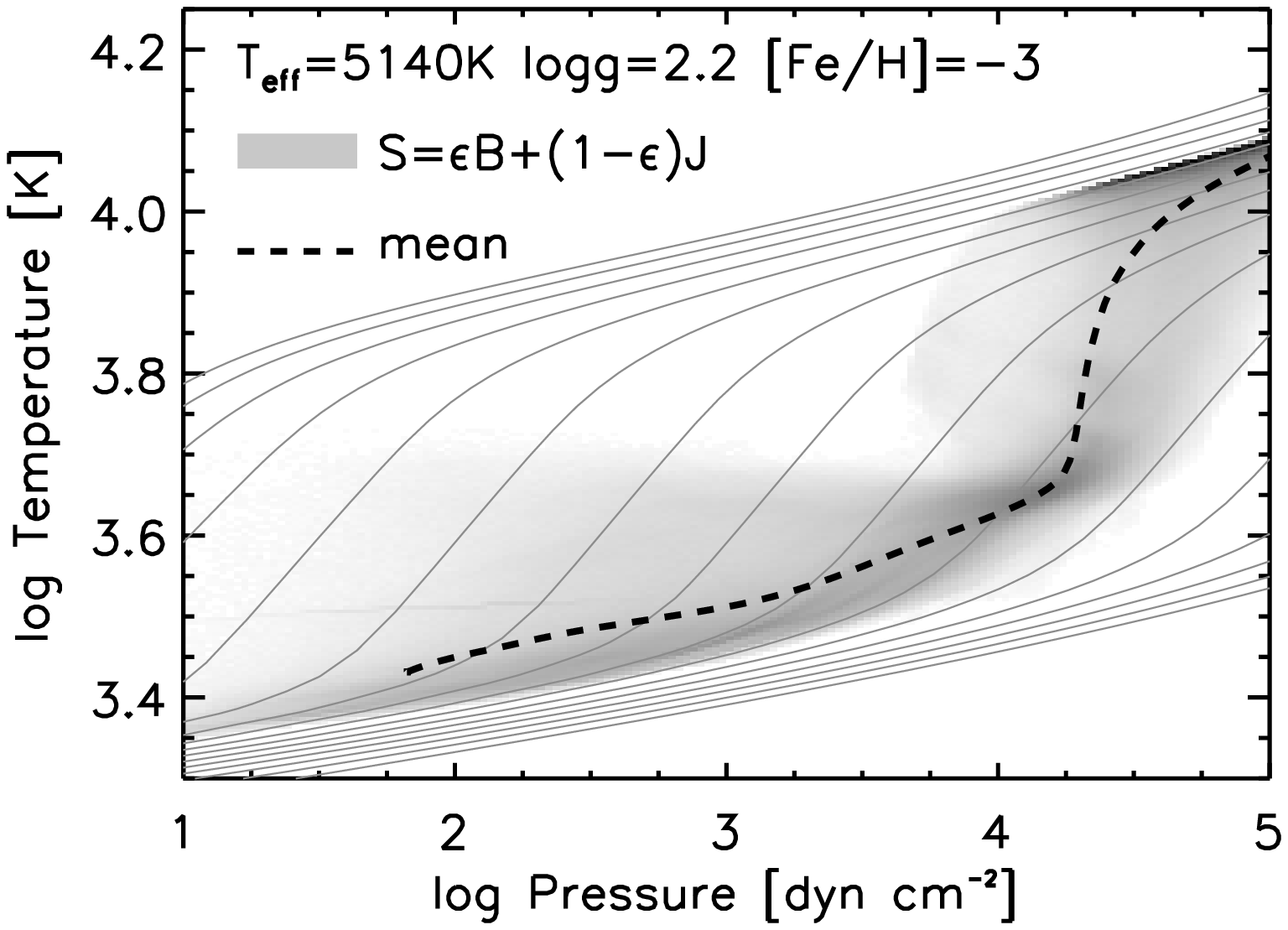} 
	\includegraphics{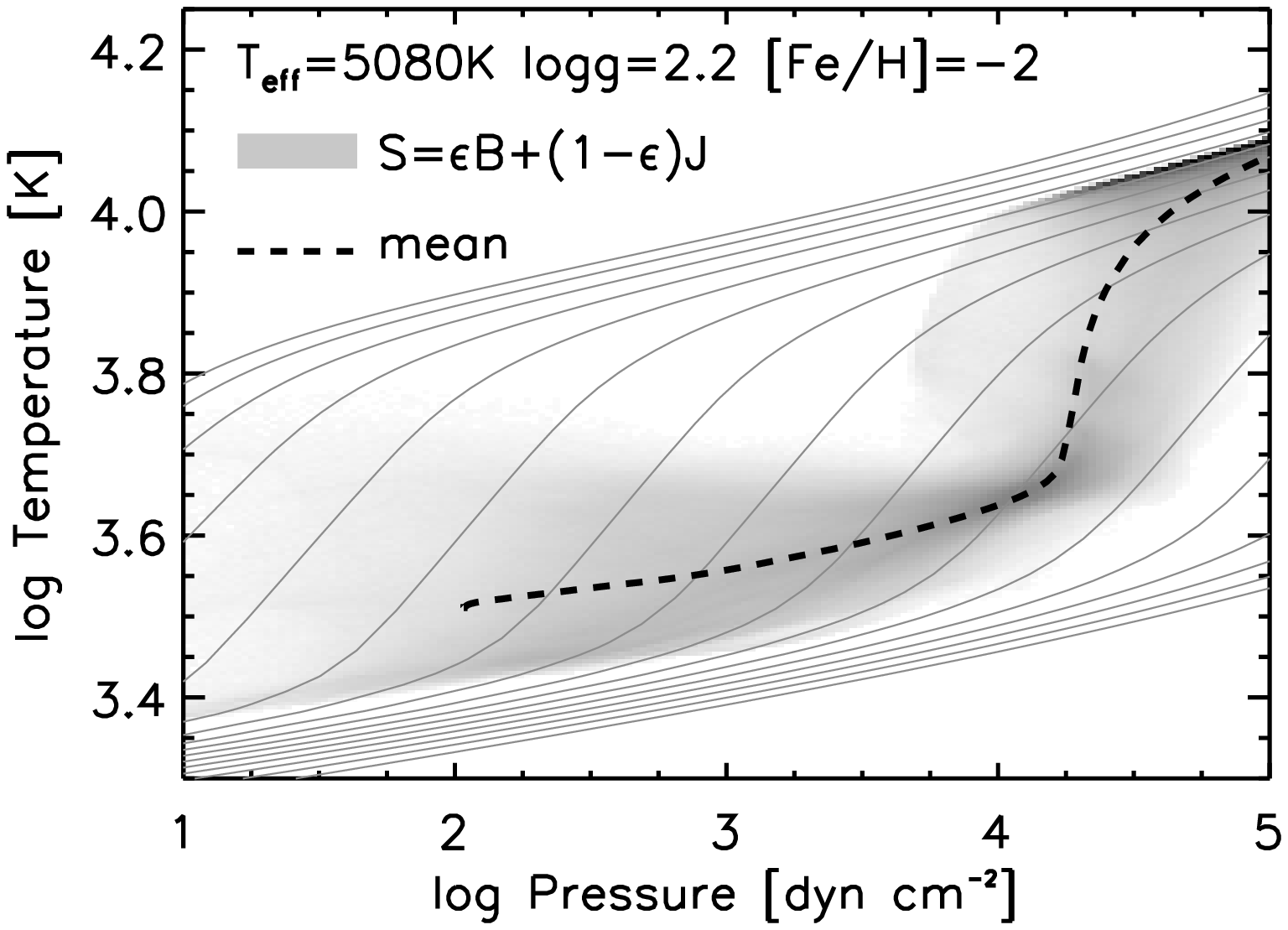} 
}
  \caption{ \emph{Grey shaded area}: distribution of temperature versus gas pressure in two representative snapshots of metal-poor red giant surface convection simulations at [Fe/H]=$-3$ (\emph{left panel}) and [Fe/H]=$-2$ (\emph{right panel}). The simulations were computed with the radiative transfer solver for the coherent scattering case. Darker shades indicate higher frequency of occurrence. 
\emph{Dashed lines}: mean stratifications, averaged over time and surfaces of constant Rosseland optical depth. \emph{Thin grey lines}: contours of constant entropy.
}
  \label{fig:entropy}
\end{figure*}

Our results indicate that assuming a Planckian source function and excluding scattering in optically thin layers is a good approximation for estimating radiative heating rates in metal-poor late-type stellar surface convection simulations.
At first sight, this is an intuitive result to some degree, since coherent scattering does not contribute directly to the energy balance in the upper photosphere: its primary effect is, in fact, to redirect photons to other directions without changing their energy. 
On the other hand, when scattering processes coexist with absorption processes in stellar atmospheres (that is, when scattering and pure absorption opacities are comparable), they affect the form of the source function to an appreciable degree and, consequently, play a crucial role in determining the actual magnitude of mean intensities and of radiative heating rates.

The non-local nature of scattering and the presence of temperature and density inhomogeneities in late-type stellar photospheres renders the problem of determining the effective feedback of scattering on the energy balance very complicated.
It is therefore necessary to carry out numerical simulations to verify our hypothesis that excluding the contribution of scattering to extinction in the optically thin layers may work as a good approximation to the correct solution.
With hindsight, we can use our results to justify why our approximation works.

Below the optical surface, the assumption of a Planckian source function is relatively accurate because of the dominance of absorption that contributes to the thermalization of the radiation field.
Proceeding upwards, the opacity changes from absorption-dominated to scattering-dominated. In metal-poor late-type stars (and particularly in giants), this transition is fairly rapid, as we have seen (Fig.~\ref{fig:epsilon1}).
In the optically thin regions of the photosphere, where scattering processes dominate, the mean intensity is to first order set by the physical conditions in the layers located underneath, near the optical surface.
By neglecting the contribution of scattering to the opacity in surface layers and assuming that the sole role of scattering there is to redirect photons, we can provide a rough estimate of the mean intensity in the upper photospheric layers and simulate these non-local effects in the radiative transfer even though we have adopted a Planckian source function everywhere for the calculations.
Naturally, the mean intensities predicted by the coherent scattering case and by the approximated treatment with no scattering contribution to extinction in the streaming regime disagree to some extent. Small differences also arise in practice from the different weighting of opacities in the calculation of the mean extinction coefficients $\chi_i^J$ and $\chi_i^{J\,\ast}$ (Eq.~\ref{eq:binstream} and~\ref{eq:chiJ1}) and can contribute to small deviations in terms of predicted radiative heating rates.

Let us now consider what happens when we start from a snapshot of the simulation assuming a Planckian source function and no scattering contribution to extinction in optically thin layers and switch to the scattering-as-absorption case. 
Changes to the total radiative heating rate in the uppermost layers are most significant in the first bin that groups the wavelengths for which the formation region is located near the optical surface, while they are largely negligible for the remaining three bins.
This is illustrated in Fig.~\ref{fig:qradmgiant}, which shows the comparison between the horizontally averaged instantaneous radiative heating rates in individual bins computed for a representative snapshot of the metal-poor red giant simulation at [Fe/H]=$-3$ that assumes no scattering contribution to extinction in the upper layers, before and after switching to the scattering-as-absorption case.
Where the contribution of scattering to extinction in the streaming regime is neglected,  the radiative heating rate in the first bin and in optically thin layers can be approximated by
\begin{equation}
Q_{\mathrm{rad},1}^{\,\ast} \, {\approx}\, 4 \pi \,  \kappa_1^J  \, ( J_1-B_1).
\label{eq:qstream1b1}
\end{equation}
The steepness of the temperature gradient near the optical surface \citep{asplund99,collet07} in the simulations where the contribution of scattering to extinction in optically thin layers is neglected ensures that the $J_1-B_1$ split is generally positive in the upper photosphere.
As we will discuss below, this leads to radiative re-heating that maintains the temperature away from a pure adiabatic stratification.
When scattering is switched to pure absorption, the opacity in the first bin significantly increases with regard to other bins, particularly in the upper photospheric layers. 
The value of the mean intensity  $J_1$ is controlled by the conditions around the optical surface, where the changes in terms of opacity are small, and remains roughly the same, so that the $J_1-B_1$ split is also unchanged to first approximation. 
The radiative heating rate in the first bin therefore becomes
\begin{equation}
Q_{\mathrm{rad},1}^{\,\ast\ast} \,\approx\, 4 \pi \,  (\kappa_1^J+\sigma_1^J) \, ( J_1-B_1) \,>\, Q_{\mathrm{rad},1}^{\,\ast} \,>\, 0,
\label{eq:qstream2b1}
\end{equation}
which explains why the temperature increases in the upper photosphere when scattering is treated as pure absorption.

\citet{hayek10} have shown that continuous scattering plays a negligible role in shaping the temperature stratification in simulations of solar-type, solar-metallicity, stars. 
However, the situation considered here is qualitatively different from theirs: in the upper photospheres of stars similar to the Sun, continuum scattering represents only a negligible fraction of the total opacity, so that excluding it or including it as true absorption in optically thin layers does not lead to significant differences in terms of the predicted temperature and density stratifications.

Throughout our analysis, we treated line opacities as pure absorption when computing the opacity bins.
\citet{hayek10} have also considered the case where line scattering is treated as such in the radiative transfer calculations, estimating line photon destruction probabilities by means of the \citet{vanregemorter62} approximation.
According to their calculations, the overall impact on the temperature stratification is fairly small, with differences of less than $40$~K for $\log\tau_\mathrm{Ross} \ga -4$ with respect to the case where line scattering is treated as pure absorption. 
It would be interesting to examine whether the effects of line scattering on heating rates are more significant for metal-poor stellar surface convection simulations, but this goes beyond the scope of the present discussion; we therefore reserve the study of these effects for future work.

Another limitation to the accuracy of the calculations is the opacity binning approximation of non-grey radiative transfer.
For the current analysis, we used a somewhat crude choice for the binning with only four opacity groups and no explicit division in the sense of wavelength.
Allowing for more than four bins and grouping opacities according not only to strength but also to wavelength would increase the accuracy of the calculations, but it is unlikely that it would affect the main conclusions;
the exact value of the difference between the temperature stratifications predicted with various treatments of scattering would depend on the details of the binning, but the deviations from the individual simulations computed with four bins would overall be relatively small.
We performed test calculations with twelve opacity bins (with bin selection based on strength and wavelength), neglecting scattering extinction in optically thin layers or including it as true absorption, and found that the main results are very similar to the ones obtained with four bins.

We also tested whether the cool stratification may be an artifact caused by the choice of mean temperature and density stratification in the definition of the mean opacities in the streaming limit (Eq.~\ref{eq:binstream} and~\ref{eq:chiJ1}). 
As mentioned in Sect.~\ref{sec:opacbin}, we used for this task the mean stratifications from relaxed simulations that we computed excluding the scattering contribution to extinction in the streaming regime.
A question is whether the computed streaming-regime bin opacities could force the temperature stratifications in the standard case to remain artificially cool -- as well as in the case where we assume a Planckian source function and no scattering contribution to extinction in optically thin layers. 
To check this, we recomputed the streaming-regime bin opacities for the metal-poor red giant at [Fe/H]=$-3$ for all three scattering treatments, instead using the hot mean stratification from the simulation in which scattering is treated as absorption everywhere. 
We found consistency with the previous results: when scattering is correctly treated, or the contribution of scattering is left out from the streaming-regime opacities, the hot stratification immediately cools down to the same level that it had in the original simulation. 
In the scattering-as-absorption case, the temperature in the upper atmospheric layers of the simulations shows some degree of sensitivity to the mean stratification adopted for the opacity binning calculations. When the mean stratification for the opacity binning calculations is iteratively adjusted to match the actual mean stratification resulting from the scattering-as-absorption simulations, the temperatures in the surface layers eventually attain lower values than predicted in first approximation (e.g., about $70$~K lower for the [Fe/H]=$-3$ red giant simulation and about $200$~K lower for the [Fe/H]=$-3$ turnoff star), although they still remain significantly higher than in the coherent scattering and no-scattering-in-streaming-regime cases.  Thus, our main conclusions are unaltered. 
We again stress that the scattering-as-absorption approximation is generally unphysical, and that the no-scattering-in-streaming-regime approximation provides a better approximation to the correct coherent scattering solution. 

Finally, Fig.~\ref{fig:entropy} shows the temperature-pressure distributions in two representative snapshots from the [Fe/H]=$-3$ and [Fe/H]=$-2$ red giant simulations with proper treatment of coherent scattering in the radiative transfer. 
Although the distributions are limited from below by the curves defined by the process of H$_2$ formation under adiabatic conditions, the stratification itself in the upper photosphere is actually not adiabatic. 
In all columns in the upper part of the simulations, the entropy gradient is generally directed outwards -- as expected for convectively stable regions -- with only a few localized  sites where the gradients are shallow or flat.
Hence, even in the coherent scattering case (and also where the contribution of scattering to extinction in optically thin layers is excluded), radiative transfer is still responsible for regulating the temperature stratification in the uppermost layers, which would otherwise be even cooler if they were solely controlled by adiabatic processes.
Consequently, we are confident that the goodness of our approximation of radiative transfer with coherent scattering is not a spurious coincidence because the photospheric stratifications are possibly too close to the adiabatic limit.

%__________________________________________________________________
\section{Concluding remarks}
We have carried out 3D radiative hydrodynamic simulations of convection at the surface of three metal-poor stars (two giants and one turnoff star) using different treatments of continuum scattering for the solution of the radiative transfer equation.
These consist of two approximated implementations in which we assume a Planckian source function, excluding or including the contribution of scattering to opacity in the optically thin layers of the photosphere, and one implementation in which we self-consistently solve the radiative transfer equation for a non-Planckian source function that includes a coherent isotropic continuum scattering term.
We found that simulations computed with the last-mentioned method and with the approximation in which we exclude scattering from the total extinction in optically thin layers produce temperature and density stratifications that agree well with each other.
This is essentially because, to first approximation, coherent isotropic scattering processes do not contribute directly to radiative energy exchanges in the upper photosphere.
In particular, both solutions predict a temperature stratification in the upper photospheric layers of metal-poor late-type stars that is lower on average than the one predicted by 1D hydrostatic model atmospheres.
This result is consistent with the findings by \citet{asplund99} and \citet{collet07}: we therefore rule out that the cool temperature stratification predicted by those works can be ascribed to their approximated treatment of scattering.

The advantage of using the approximation of a Planckian source function and no continuum scattering contribution to extinction in optically thin layers is that it allows us to achieve a sufficiently accurate representation of the radiative heating rates in the upper photosphere and produces temperature density stratifications very similar to the one obtained with the proper, iterative, solution, but at a much lower computational expense.
In our analysis, we also showed that assuming a Planckian source function and treating scattering as true absorption everywhere in the simulation -- as in present {\sc CO$^5$BOLD} models -- leads to systematically hotter temperature stratifications.
At metallicities of [Fe/H]=$-3$ and [Fe/H]=$-2$, this approximation can produce deviations from the correct solution of the order of a few hundred kelvins in the upper photosphere, which can significantly affect the predictions from spectral-line-formation calculations.

\bibliographystyle{aa} 
%%\bibliography{rcollet}

\end{document}